\documentclass[sigconf]{acmart}

\usepackage{tikz}
\usepackage[most]{tcolorbox}
\usepackage{gensymb}
\usepackage{todonotes}
\usepackage{amsmath,amsfonts}
\usepackage{algorithm}
\usepackage[noend]{algpseudocode}
\usepackage[symbol]{footmisc}
\usepackage{tikz}
\usepackage{amsmath}
\usepackage{multirow}
\usepackage{booktabs}

\newcommand*\circled[1]{\tikz[baseline=(char.base)]{
            \node[shape=circle,fill,inner sep=1pt] (char) {\textcolor{white}{#1}};}}
\usepackage{xcolor}
\usepackage{pgfplots}
\usepackage{tikz}
\pgfplotsset{compat=1.8}
\pgfplotsset{
    width=\textwidth,
}
\usepackage{lipsum}
\usepackage{pgfplotstable}
\usetikzlibrary{pgfplots.groupplots}

\usepackage{tikz}

\definecolor{codegreen}{rgb}{0,0.6,0}
\definecolor{codegray}{rgb}{0.5,0.5,0.5}
\definecolor{codepurple}{rgb}{0.58,0,0.82}
\definecolor{mGreen}{rgb}{0,0.6,0}
\definecolor{mGray}{rgb}{0.5,0.5,0.5}
\definecolor{mPurple}{rgb}{0.58,0,0.82}
\definecolor{backcolour}{rgb}{0.95,0.95,0.92}

\definecolor{RYB1}{RGB}{80, 99, 42}
\definecolor{RYB2}{RGB}{215, 227, 191}
\definecolor{RYB3}{RGB}{198, 187, 174}
\definecolor{RYB4}{RGB}{146, 205, 220}
\definecolor{RYB5}{RGB}{238, 144, 34}
\definecolor{RYB6}{RGB}{142, 172, 59}

\pgfplotscreateplotcyclelist{colorbrewer-RYB}{
{RYB1!50!black,fill=RYB1},
{RYB2!50!black,fill=RYB2},
{RYB3!50!black,fill=RYB3},
{RYB4!50!black,fill=RYB4},
{RYB5!50!black,fill=RYB5},
{RYB6!50!black,fill=RYB6},
}

\usepackage{lscape} 
\usepackage{pdflscape}
\usepackage{afterpage}
\usepackage{capt-of}
\usepackage{listings}
\usepackage{float}
\usepackage{array,multirow,graphicx}
\usepackage[titletoc]{appendix}
\usepackage{caption}
\usepackage{subcaption}
\usepackage{soul}
\usepackage{pifont}

\definecolor{ggreen}{HTML}{2CC225}
\definecolor{yyellow}{HTML}{C2C80A}
\definecolor{bbrown}{HTML}{8e4603}

\definecolor{codegreen}{rgb}{0,0.6,0}
\definecolor{codegray}{rgb}{0.5,0.5,0.5}
\definecolor{codepurple}{rgb}{0.58,0,0.82}
\definecolor{mGreen}{rgb}{0,0.6,0}
\definecolor{mGray}{rgb}{0.5,0.5,0.5}
\definecolor{mPurple}{rgb}{0.58,0,0.82}
\definecolor{backcolour}{rgb}{0.95,0.95,0.92}

\lstdefinestyle{CStyle}{
    commentstyle=\color{mGreen},
    keywordstyle=\color{magenta},
    numberstyle=\tiny\color{mGray},
    stringstyle=\color{mPurple},
    basicstyle=\sffamily\footnotesize,
    frame=lrtb,
    breakatwhitespace=false,         
    breaklines=true,                 
    captionpos=b,                    
    keepspaces=true,                 
    numbers=left,                    
    numbersep=5pt,                  
    showspaces=false,                
    showstringspaces=false,
    showtabs=false,                  
    tabsize=2,
    language=C
}

\lstdefinestyle{CStyle1}{
    commentstyle=\color{mGreen},
    keywordstyle=\color{magenta},
    numberstyle=\tiny\color{mGray},
    stringstyle=\color{mPurple},
    basicstyle=\sffamily\footnotesize,    frame=lrtb,
    breakatwhitespace=false,         
    breaklines=true,                 
    captionpos=b,                    
    keepspaces=true,                 
    numbers=left,                    
    numbersep=5pt,                  
    showspaces=false,                
    showstringspaces=false,
    showtabs=false,                  
    tabsize=2,
    language=C
}

\lstdefinestyle{mystyle}{
    commentstyle=\color{codegreen},
    keywordstyle=\color{magenta},
    numberstyle=\tiny\color{codegray},
    stringstyle=\color{codepurple},
    basicstyle=\sffamily\footnotesize,
    breakatwhitespace=false,         
    breaklines=true,                 
    captionpos=b,                    
    keepspaces=true,                 
    numbers=left,                    
    numbersep=5pt,                  
    showspaces=false,                
    showstringspaces=false,
    showtabs=false,                  
    tabsize=2,
    language=C
}
\lstdefinestyle{trans}{
    commentstyle=\color{codegray},
    numberstyle=\tiny\color{codegray},
    stringstyle=\color{codepurple},
     basicstyle=\sffamily\footnotesize,
    frame=lrtb,
    breakatwhitespace=false,         
    breaklines=true,                 
    captionpos=b,                    
    keepspaces=true,                 
    numbers=left,                    
    numbersep=5pt,                  
    showspaces=false,                
    showstringspaces=false,
    showtabs=false,                  
    tabsize=2,
     language=[x86masm]Assembler,  escapeinside={\%*}{*)},   
     }     

\AtBeginDocument{%
  \providecommand\BibTeX{{%
    \normalfont B\kern-0.5em{\scshape i\kern-0.25em b}\kern-0.8em\TeX}}}

\copyrightyear{2023}
\acmYear{2023}
\setcopyright{acmlicensed}\acmConference[CCS '23]{Proceedings of the 2023 ACM SIGSAC Conference on Computer and Communications Security}{November 26--30, 2023}{Copenhagen, Denmark}
\acmBooktitle{Proceedings of the 2023 ACM SIGSAC Conference on Computer and Communications Security (CCS '23), November 26--30, 2023, Copenhagen, Denmark}
\acmPrice{15.00}
\acmDOI{10.1145/3576915.3623092}
\acmISBN{979-8-4007-0050-7/23/11}

\begin{document}


\title{LeakyOhm: Secret Bits Extraction using Impedance Analysis}

\author{Saleh Khalaj Monfared}
\affiliation{%
  \institution{Worcester Polytechnic Institute}
  \city{Worcester}
  \state{MA}
  \country{USA}
}
\email{skmonfared@wpi.edu}

\author{Tahoura Mosavirik}
\affiliation{%
  \institution{Worcester Polytechnic Institute}
 \city{Worcester}
  \state{MA}
  \country{USA}
 }
\email{tmosavirik@wpi.edu}

\author{Shahin Tajik}
\affiliation{%
  \institution{Worcester Polytechnic Institute}
 \city{Worcester}
  \state{MA}
  \country{USA}
  }
\email{stajik@wpi.edu}

\renewcommand{\shortauthors}{Saleh Khalaj Monfared, Tahoura Mosavirik, \& Shahin Tajik}
\begin{abstract}
The threats of physical side-channel attacks and their countermeasures have been widely researched.
Most physical side-channel attacks rely on the unavoidable influence of computation or storage on current consumption or voltage drop on a chip.
Such data-dependent influence can be exploited by, for instance, power or electromagnetic analysis.
In this work, we introduce a novel non-invasive physical side-channel attack, which exploits the data-dependent changes in the impedance of the chip.
Our attack relies on the fact that the temporarily stored contents in registers alter the physical characteristics of the circuit, which results in changes in the die's impedance. 
To sense such impedance variations, we deploy a well-known RF/microwave method called scattering parameter analysis, in which we inject sine wave signals with high frequencies into the system's power distribution network (PDN) and measure the echo of the signals. 
We demonstrate that according to the content bits and physical location of a register, the reflected signal is modulated differently at various frequency points enabling the simultaneous and independent probing of individual registers.
Such side-channel leakage challenges the $t$-probing security model assumption used in masking, which is a prominent side-channel countermeasure.
To validate our claims, we mount non-profiled and profiled impedance analysis attacks on hardware implementations of unprotected and high-order masked AES.
We show that in the case of the profiled attack, only a single trace is required to recover the secret key.
Finally, we discuss how a specific class of hiding countermeasures might be effective against impedance leakage.

\end{abstract}




\settopmatter{printfolios=true}
 \maketitle

\section{Introduction}\label{sec:introduction}

Physical side-channel leakages can compromise the security of cryptographic implementations on integrated circuits (ICs). Such leakages exist due to the inevitable impact of computation and storage on current consumption or voltage drop on a chip. These data-dependent fluctuations reveal themselves through various measurable quantities, such as power consumption~\cite{kocher1996timing}, electromagnetic emanation~\cite{kocher1999differential}, acoustic waves~\cite{genkin2017acoustic}, photon emission~\cite{ferrigno2008aes}, and thermal radiation~\cite{hutter2014temperature}. Over the last decades, these quantities have been exploited in different classes of side-channel analysis (SCA) attacks for breaking the security of various cryptographic implementations. At the same time, various countermeasures (e.g., hiding and masking) have been developed to defeat these attacks.

While current and voltage alterations have been considered the root cause of side-channel leakages, the data-dependent variation of the parameter relating current and voltage to each other via Ohm’s law, i.e., impedance, has always been ignored. The primary assumption has been that impedance is a constant parameter that is determined by the materials used in the fabrication of the PCB, chip’s die, and package. Hence, it is defined by the physical structure and size of the chip rather than the running computation on a system or stored content on a chip. For instance, adding/removing a circuit to/from a chip can cause changes in the impedance of the die. Such changes have been the basis of some hardware Trojan and tamper detection methods (on both chips~\cite{Backscattering_HT_2019,Backscattering2020Nguyen,mosavirik2023silicon} and PCBs~\cite{PDNPulse,mosavirik2023impedanceverif,mosavirik2022scatterverif}), where the malicious circuits modify the impedance of the system and, thus, can be detected. However, the effect of the circuit state or content of memory elements inside the chip on information leakage through the die’s impedance has not been studied so far. 

The contribution of impedance to side-channel leakage could be implicitly observed in a specific class of SCA attacks, namely static power analysis~\cite{moradi2014side}. For this attack, the adversary halts the circuit and exploits the data-dependent static current consumption of transistors in steady states. It was shown that the state of flip-flops on a chip leads to static current variations, leading to successful key recovery using differential power analysis (DPA). The fluctuation in the static current is indeed caused by changes in the overall impedance of the die; however, the focus in static SCA attacks~\cite{moradi2014side,moos_static_2019,cassiers2023prime} has been on the measurable quantity, i.e., the static current and the role of the impedance has never been discussed. 

Driven by the fact that impedance is also affected by the stored content on a chip, the following research questions arise: \emph{(1) Is it possible to measure the information leakage through the impedance directly? (2) What would be the consequence of impedance analysis for prominent side-channel countermeasures, i.e., masking?}

\begin{figure*}[t!]
           \centering
        \begin{subfigure}{.69\linewidth}
         \includegraphics [width=\textwidth]{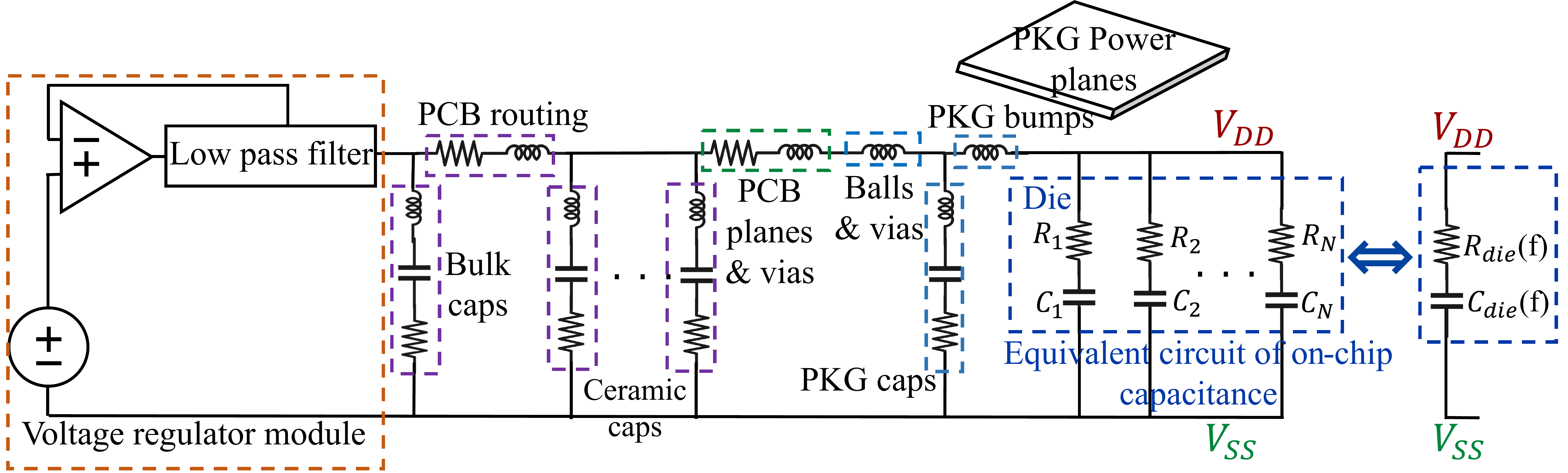}
                \caption{}
               \label{subfig:RLC_chip}
        \end{subfigure}
    \hspace{0.5pt}
        \begin{subfigure}{.28\linewidth}
         \includegraphics [width=\textwidth]{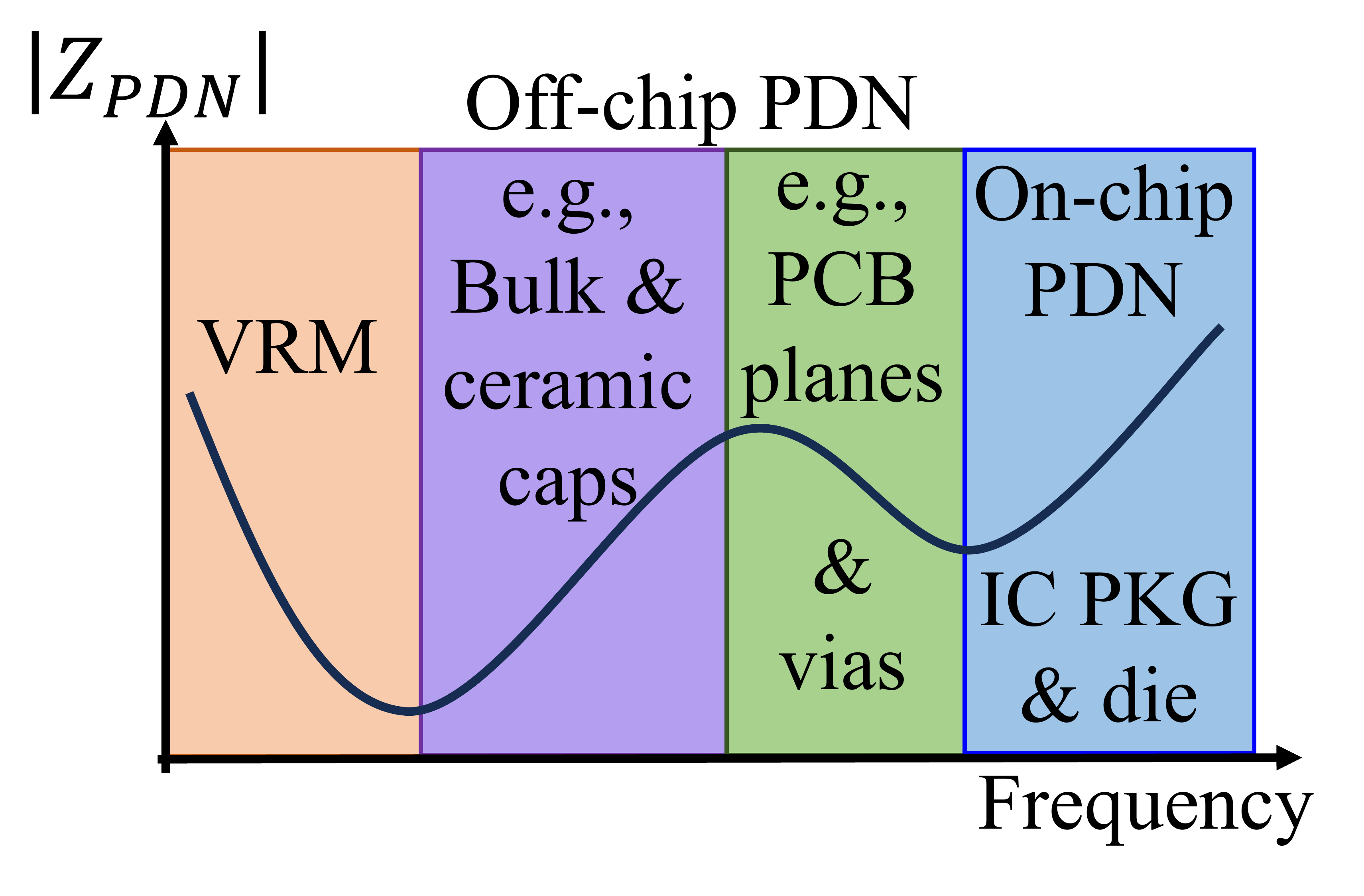}
                \caption{}
               \label{subfig:freq_disp}
                 \end{subfigure}
	\caption{(a) Equivalent RLC circuit model of the power distribution network of the PCB and chip~\cite{Bogatin-Smith-PI}. (b) Contribution of different parts of the PDN to the impedance over frequency.}
   	\label{fig:RLC_Model_freq_disp}
 \end{figure*}
 
\noindent\textbf{Our Contribution.} To answer the above questions, we present a novel non-invasive SCA attack based on directly characterizing the chip’s impedance. Our method relies on a known RF/microwave impedance characterization technique called scattering parameter analysis, in which we inject electrical sine waves with different frequencies into the power delivery network (PDN) of the chip and measure the echoes of the signals. We demonstrate that the reflected signals are modulated uniquely at various frequency points based on the register contents and physical location of a register on the chip. The transistor’s imperfections, asymmetric logic gates, and interconnects with various lengths contribute to impedance variations and, consequently, to the modulation of the reflected signals. We show that analyzing these echoes at different frequencies enables the adversary to profile register contents and read them out simultaneously in the attack phase. Therefore, such analyses provide a large number of virtual probes for different locations of the chip in the frequency domain during a time period, which challenges the central underlying assumption of the $t$-probing security model for masking schemes. To validate our claims, we launch an impedance-based SCA attack on unmasked and masked AES designs implemented on a Field Programmable Gate Array (FPGA) manufactured with a 28 nm technology. As a result, we successfully break the security of the targeted implementations by recovering their keys.


\noindent{\textbf{Remark.} The primary contribution of this work is the introduction of a new physical side-channel based on impedance analysis. A comparative experimental study between impedance analysis and other side-channels, specifically power side-channels, requires a unified setup\cite{del2015side} and deserves a future study.}




\section{Technical Background}\label{sec:Background}

\subsection{Power distribution network (PDN)}\label{Chip PDN}

The PDN is responsible for delivering low noise and constant voltage supply to the  electronic components on the PCB, from the voltage regulator module (VRM) to the power rails on the chip. 
Each component has a distinct contribution to the physical signature of the PDN at different frequency regimes.
The system's PDN is represented by an equivalent circuit model shown in Figure~\ref{subfig:RLC_chip}. 
The PDN comprises both off-chip and on-chip components, including bulk capacitors, PCB routing, ceramic capacitors, PCB planes, vias, package bumps, on-chip power planes, and transistor capacitance. 
The impedance contribution of these components to the overall PDN's impedance is different at various frequency bands.
The voltage regulator's and off-chip components' impedance dominate the PDN's impedance at lower frequencies, while on-chip components contribute mostly to the impedance at higher frequencies, as shown in Figure~\ref{subfig:freq_disp}. 
The parasitic inductance present on each capacitor is the primary cause of this impedance behavior. At high frequencies, an ideal capacitor behaves like a short circuit. 
However, the parasitic inductance on the capacitor's metals results in resonance at a particular frequency, causing it to become an open circuit at very high frequencies. 
Smaller capacitors have less parasitic inductance and resonate at higher frequencies. 
As a result, as the frequency increases, all capacitors, from large to small, become open circuits and have less impact on the PDN impedance. 
The PDN impedance at higher frequencies is dominated by the on-chip structures due to their smaller dimensions, as shown in Figure~\ref{subfig:freq_disp}.

The dashed blue region in Figure~\ref{subfig:RLC_chip} shows the equivalent RC model of the on-chip capacitance.
To model the wideband on-chip behavior of the circuit, multiple narrowband parallel RC circuits (N in total) are connected to $V_{DD}$ and $V_{ss}$. 
The succeeding subsection provides further details on the origins of on-chip PDN impedance.
 
\subsection{Sources of On-die Impedance}\label{inverter}

On-die capacitance $C_{die}$ and resistance $R_{die}$ are the dominant features of the on-chip impedance in high-frequency bands~\cite{DesignCon}.
The ranges of such frequency bands are determined based on the chip's technology and size.
Here, we explain the sources of on-die capacitance using the physical structure of a CMOS inverter.
Figure~\ref{fig:SourceOfchange} shows the cross-sectional view of an inverter, metal power grid layers, and the locations of the corresponding on-die capacitors.
According to Figure~\ref{fig:SourceOfchange}, an inverter comprises a PMOS and an NMOS transistor.
These transistors serve as switches, with the NMOS having an infinite off-resistance and a finite on-resistance. 
Meanwhile, the PMOS has a positively doped source, drain, and gate region in the form of an n-well. 
The On-die capacitance $C_{die}$ is affected by several elements, including the metal layers grid network, non-switching gate, and p-n diode junction diffusion~\cite{DesignCon}. 
Resistance in the power net, transistor channel, transistor gate, and contacts of n-well and P-substrate contribute to $R_{die}$~\cite{PowerGrid}.

\begin{figure}[t!]
   \centering \noindent
   \includegraphics[width=\linewidth]{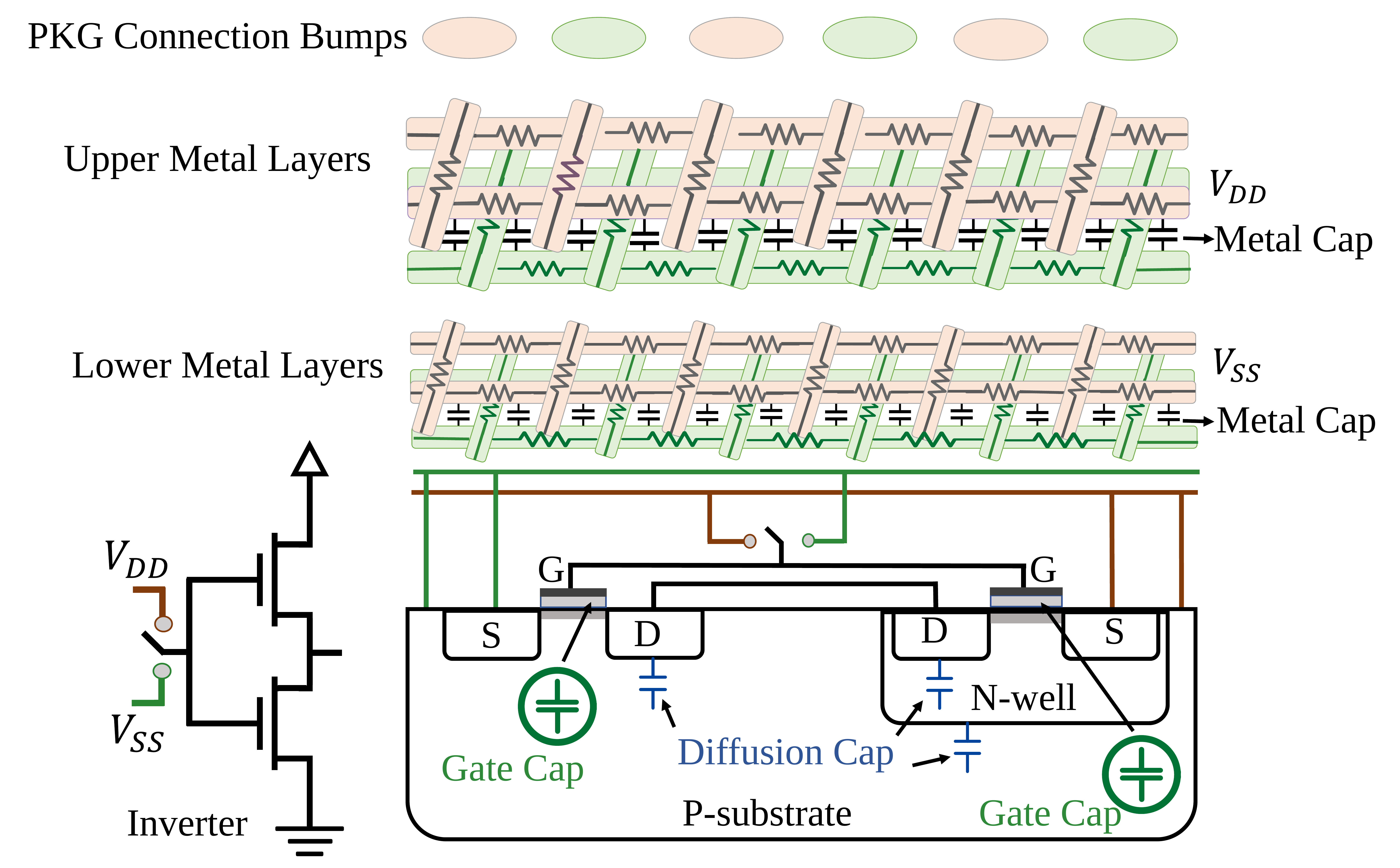}
               \label{fig:SourceOfchange}
    \caption{The physical representation of a CMOS inverter cross section and the locations of different types of on-die capacitors~\cite{mosavirik2023silicon}. The black capacitors show the capacitance of metal lines, the blue ones show the p-n diode junction diffusion capacitance, and the capacitance shown in green color corresponds to non-switching gate capacitance.}
   	\label{fig:SourceOfchange}
 \end{figure}

The location of each capacitance that contributes to $C_{die}$ is shown in Figure~\ref{fig:SourceOfchange} using different colors. 
The black color represents the metal capacitance, $C_{m}$, which pertains to the power/ground metallization grid network located on the die. 
The size of $C_{m}$ is affected by the density of the grid network, the width and distance of metal layers, and the permittivity of materials. 
Typically, $C_{m}$ is larger in upper metal layers due to denser power and ground meshes, while it is slightly smaller in lower metal layers because the power traces are less dense and thinner.
The blue color corresponds to the diffusion capacitance, $C_{d}$, which relates to the p-n diode junctions. 
It is essential to note that $C_{d}$ and $C_{m}$ only contribute to a small portion of the total $C_{die}$, while the non-switching gate capacitance, $C_{g}$, is the main contributor.

 
On the chip's PDN, all non-switching and powered-on circuits contribute to $C_{g}$. 
This is because when a transistor is powered on, it has a channel underneath the gate, contributing to $C_{die}$. 
On the other hand, when a transistor is powered off, its channel is inactive and does not significantly contribute to on-die capacitance.
Initially, when the device is not powered on, the decoupling capacitance effect of the gates is negligible. However, when the device is turned on, the channels start to form, and as a result, $C_{g}$ becomes the dominant contributor to $C_{die}$.
If the chip's design is modified, different parts of $C_{die}$ (particularly $C_{g}$) would change based on the size, location, and nature of the tamper event. This modification changes the equivalent circuit of the on-chip PDN and affects the measured signatures from the chip.


\subsection{Non-invasive Impedance Characterization}\label{S_Z_Params_General}

To characterize the impedance of the PDN in different frequencies, S (Scattering) or Z (Impedance) parameters are deployed~\cite{bogatin2010signal,pupalaikis2020s}.
Every circuit/electronic component can be described as a one or multi-port network.
S parameters directly represent the attenuation, reflection transmission ratio of the signal at
each port of such network over the frequency domain to the applied electromagnetic field~\cite{pozar2011microwave}. For instance, $S_{11}$ quantifies how much power is reflected from the the port 1 if the EM field is transmitted from port 1.
In frequency domain analysis, waveforms are represented by sine waves. 
Frequency, amplitude, and phase are the three terms that can fully characterize a sine wave.
Thus, we utilize both the amplitude and phase response in the frequency domain to accurately characterize the chip at each frequency point.
A Vector Network Analyzer (VNA) is an instrument that measures the transmitted and/or reflected power of a signal that goes into and comes back from a component. 
We use a VNA to inject sine waves into the chip at every frequency point to record the chip's PDN's reflected response .
The impedance profile can be easily derived from the reflection coefficient.
Equation~\ref{Conversion_to_Z} expresses the relationship between the input impedance of the device under test (DUT) and the reflection coefficient:

\begin{equation}\label{Conversion_to_Z}
Z_{DUT}=Z_0\dfrac{1+S_{11}}{1-S_{11}},
\end{equation}

where $S_{11}$ is the reflection coefficient, $Z_0$ represents
the reference impedance of the VNA which is $50~\Omega$, and $Z_{DUT}$ corresponds to the impedance obtained from $S_{11}$. 
We only deploy $S_{11}$ in our proposed method as the VNA can directly measure it from the chip.
However, based on Equation~\ref{Conversion_to_Z}, it is observable that the reflection coefficient is another representation of the impedance.



\begin{figure}[t]
   \centering \noindent
   \includegraphics[width=\linewidth]{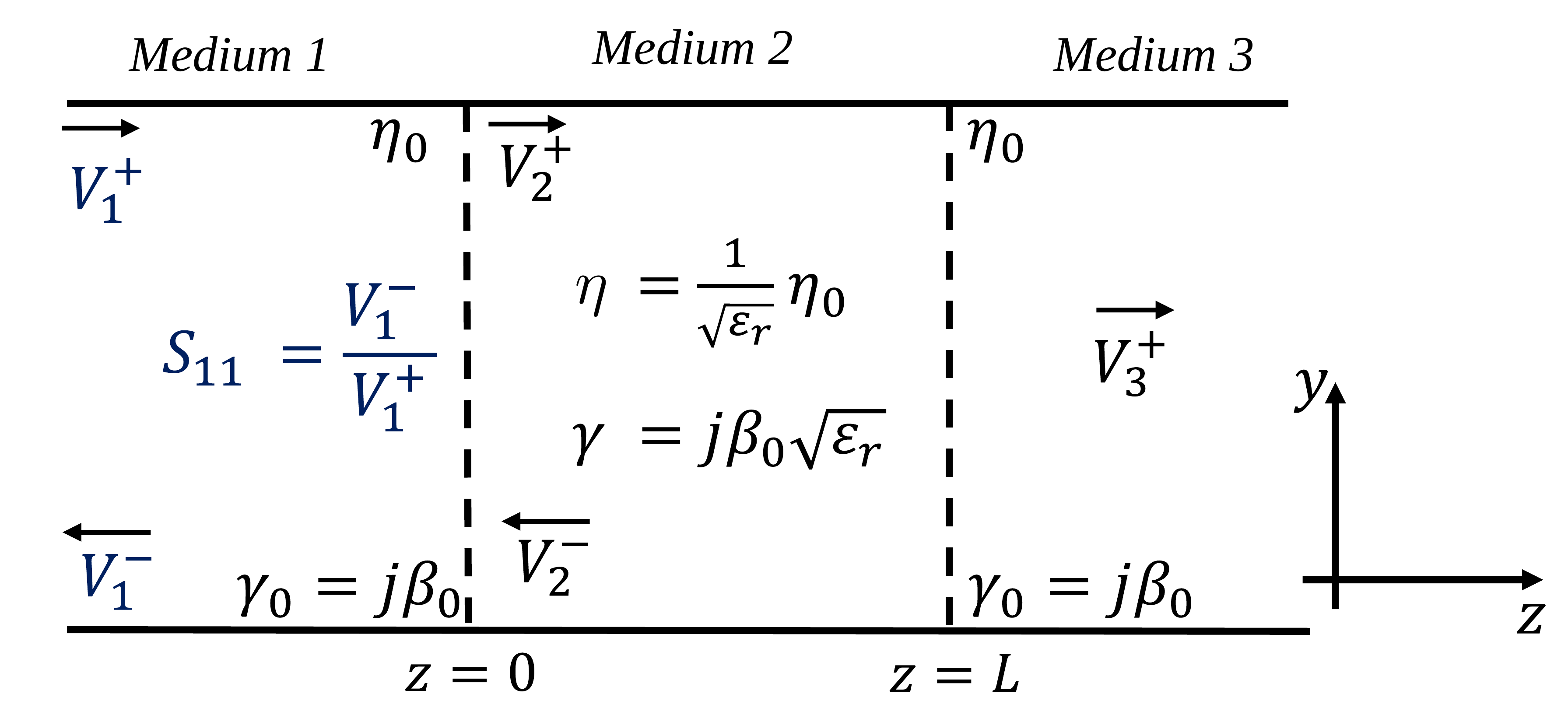}
	\caption{The simplified (ideal) transmission line model for normal uniform plane wave incidence on different media (the characteristic impedance of medium 2  is different from medium 1 and 3)~\cite{mosavirik2023silicon}.}
   	\label{fig:S_param_TL_ideal}
 \end{figure} 

  \begin{figure*}[!t] 
\centering
\includegraphics[width=0.95\linewidth]{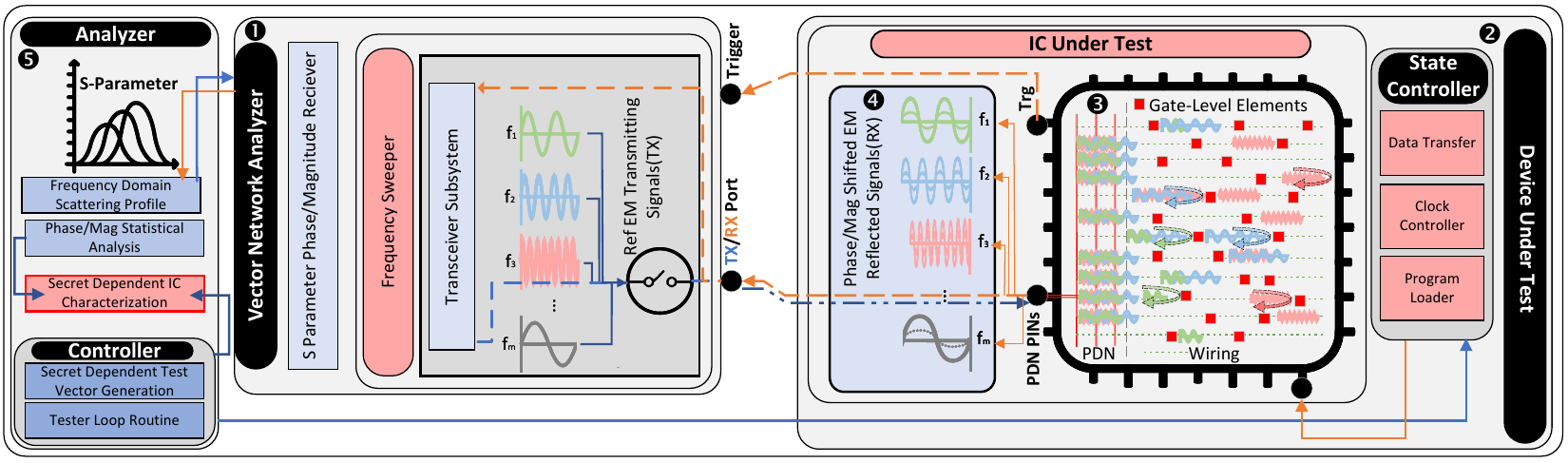}
\caption{High-level overview of \textsc{LeakyOhm}'s methodology and attack flow}
\label{overview}
\end{figure*}

We further explain the changes that occur to the injected voltage wave by the VNA into the chip by analyzing the ideal transmission line model.
This model is the backbone of more complex circuits, and understanding its theoretical foundation clarifies our methodology's mechanism.
Figure~\ref{fig:S_param_TL_ideal} shows an ideal transmission line model where there is a change in the characteristic impedance and propagation constant of medium 2 that are represented by $\eta$ and $\gamma$, respectively. 
For simplicity, we assume that medium 1 and medium 3 are lossless, thus giving a characteristic impedance of $\eta_0$ and a corresponding propagation constant of $\gamma_0=j\beta_0$. 
We consider medium 2 a non-magnetic ($\mu_r=1$) medium with a relative permittivity of $\varepsilon_r$. Noting that $\varepsilon_0$ and $\mu_0$ are the permittivity and permeability of the free space, respectively, and $\beta_0$ denotes the free space wave number, we can 
consider $\beta_0 = 2\pi f\sqrt{\varepsilon_0\mu_0}$ and rewrite the second medium's propagation constant as $\gamma=j\beta_0\sqrt{\varepsilon_r}$.
Considering $\eta_0=\sqrt{\mu_0/\varepsilon_0}$, we can rewrite the characteristic impedance of medium 2 as $\eta=\sqrt{1/\varepsilon_r}\eta_0$.
The VNA injects a voltage wave with the known amplitude of $V_1^+$ in medium 1, and the reflected voltage wave has an amplitude of $V_1^-$.
After $V_1^+$ is injected, multiple reflections and transmissions occur in the lines.
Based on the model in Figure~\ref{fig:S_param_TL_ideal}, the lines’ voltages can be written as~\cite{pozar2011microwave}: $V_1(z)= V_1^+ e^{-j\beta_0z}+V_1^- e^{+j\beta_0z}$, $V_2(z)= V_2^+ e^{-\gamma z}+V_2^- e^{+\gamma z}$, and $V_3(z)= V_3^+ e^{-j\beta_0z}$, where ${V_i^+}$ and ${V_i^-}$ ($i=1,2,3$) are forward and backward voltage waves through/from the medium $i$; however, we assume that there exists no backward voltage wave in medium 3, for simplicity. 

$V_1^+$ is a known parameter (injected by VNA), whereas $V_1^-$, $V_2^+$, $V_2^-$, and $V_3^+$ are unknown values.
We apply the boundary conditions on the voltage wave components at the interfaces of the media and find all these four unknowns.
We are interested in obtaining the ${S_{11}}$ in medium 1 which can be derived as
\begin{equation}\label{TL_equations}
S_{11}(f,\varepsilon_r,L)= {{\dfrac{V_1^-}{V_1^+}}} =\dfrac{(\eta^2-\eta_0^2)({1-e^{+2j\gamma L})}}{{(\eta_0+\eta)^2 - (\eta-\eta_0)^2{e^{+2j\gamma L}}}}
\end{equation}
where $L$ is the length of the path that the injected wave voltage travels.
From Equation~\ref{TL_equations}, it can be concluded that the reflection coefficient depends on three parameters: the frequency band of interest, the relative permittivity of the sample, and the length of the wave's traveling path.
On the other hand, the dependence of ${S_{11}}$ on the frequency has another aspect: frequency and wavelength are inversely proportional to each other.
This explains why we can detect smaller size changes in the chip's configuration at higher frequencies.
When registers with different placement and routing are exposed to the incident wave injected from the VNA, the changes occurring in Equation~\ref{TL_equations} parameters will result in a change in the ${S_{11}}$ profile at distinct frequencies.
For example, when the placement and routing of the circuit is altered, L is changed, and this would cause the chip's reflection response to be different for different placements and routings.


\subsection{Masking and t-Probing Model}
Masking is the prominent countermeasure against SCA attacks due to its sound theoretical and mathematical foundations. 
In masking schemes, the computation on a chip is distributed between a couple of shares (multi-party computation) and the intermediate computations dealing with the secrets (secret sharing).
The number of shares defines the order of the masking and its resiliency against SCA and probing attacks. 
For cryptographic implementations, the key and plaintext should be represented in a shared form, and the entire computations are performed on shares. At the end of the computation, the ciphertext should be obtained by recombining the output shares.
The main advantage of masking is that it can be evaluated in formal security models. 
For instance, in Boolean masking schemes, every random bit $x$ is represented by $(x_0,\ldots,x_d)$ in such a way that $x=x_0 \oplus \ldots \oplus x_d$. 
According to~\cite{chari1999towards}, an adversary who is limited to the $d$\textsuperscript{th} order SCA or the number of probes can be defeated by a secret sharing with $d+1$ shares. Moreover, it was shown that measurements of each share $x_i$ are adversely influenced by Gaussian noise, and thus, the number of noisy traces needed to extract $x$ grows exponentially with the number of shares~\cite{prouff2009statistical}.

On the other hand, the $t$-probing model, which was first introduced in the seminal work of Ishai et al.~\cite{ishai2003private} can be deployed for the security analysis of masking schemes.
In this model, it is assumed that the attacker is limited to at most $t$ physical probes to observe the computation on wires of the circuit at each time period (e.g., one clock cycle).
In such a case, at least $t+1$ shares are needed to prevent the attacker from learning any sensitive information from $t$ observations.
It has been demonstrated that the two aforementioned leakage models are related by reducing the security in one model to the security of the other to unify the leakage models and so simplify the analysis of SCA countermeasures~\cite{duc2019unifying}.
In other words, placing $t=d$ physical probes on the wires of the target circuit is equivalent of $d$\textsuperscript{th}-order noisy SCA attack.
While there have been some sophisticated SCA attacks violating this assumptions~\cite{krachenfels2021real}, for most practical SCA attacks, it is reasonable to assume that the adversary has a limited number of probes due to practical issues due to, for instance, the lack of spatial space~\cite{specht2018dividing,kleindiek} to accommodate several physical probes or increased noise in the case of higher-order power analysis.
As a result, several constructions, security proofs, and multiple implementations have been reported.



 Note that, in several countries, protection against SCA attacks is one of the criteria defined by certification bodies. Among various SCA countermeasures, masking schemes have been widely in use for more than a decade in many secure ICs, such as smartcard chips.

\section{System-Level Impedance Analysis}\label{sec:sesytemlevel}

In this section, we propose and describe a systematic approach to analyze and interpret impedance profiles as new side-channel leakage criteria for integrated circuits. 

\subsection{High-Level Representation}
Although the nature of the back-scatter analysis is entirely non-invasive, the proposed SCA involves active measurements. Specifically, as indicated earlier, a series of test electromagnetic signals are generated in a tester device (e.g., VNA) and transmitted through the output port directly to the DUT's PDN. Consequently, the signals are collected as input at the tester's receiver, which can then be used to characterize the impedance of the DUT. Figure \ref{overview} depicts a high-level overview of the workflow, setup, and approach of our attack.

As indicated, in \circled{1}, VNA generates the test signals over a particular frequency band by employing an embedded programmable \textit{Frequency Sweeper}. During the experiments, these signals are selected in the \textit{Transceiver Subsystem} and are sent through the \textit{TX} port of the VNA. 
At the same time in \circled{2}, DUT is equipped with a \textit{State Controller} that preserves the target IC in a specific state.
\textit{TX} signals are received at IC's PDN interface. As shown in \circled{3}, based on the frequency of the transmitted signal, each test signal propagates differently throughout the chip's PDN, and hence, different propagation behavior of each of them yields in certain \textit{Magnitude} and \textit{Phase} when they are reflected back to the PDN pins (\circled{4}).
Then, VNA detects and measures the reflected signal (\textit{RX}) at each frequency and estimates scattering parameters (S-Parameters) as the final measurement. In \circled{5}, measurement data are sent to the analyzer for characterization. Analyzer performs an iterative profiling procedure to exploit the variations of reflected \textit{RX} signals in both \textit{Magnitude} and \textit{Phase} at each selected frequency (\textit{$f(|S_{11}|,\angle S_{11})$}). 

\subsection{Systematic Impedance Analysis}
In order to apply the proposed attack as an end-to-end SCA, here we present an abstract and systematic definition of impedance analysis. Inspired by telecommunication methodologies and terminologies, we can consider a communication system to describe the workflow of our attack.

\begin{figure}[t!]
  \centering \noindent
   \includegraphics[width=.6\linewidth]{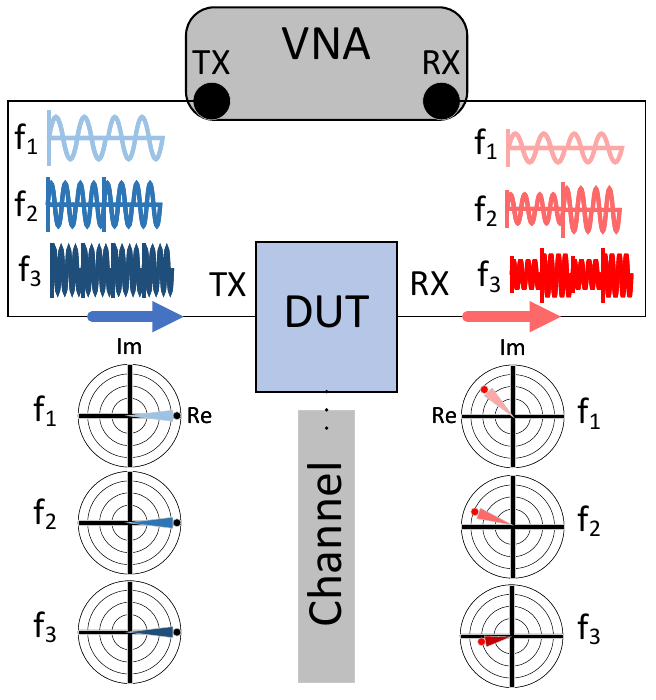}
	\caption{Systematic representation of impedance analysis inspired by communication systems.}
            \label{fig:systematic}                 
 \end{figure}

At the high-level, test signals at different frequencies, which are generated by the VNA, could be treated as data \textit{carries}. The DUT is the noisy communication \textit{channel}, and the received signals are the measurements that are to be analyzed and decoded. As depicted in Figure \ref{fig:systematic}, a series of signals on pre-determined frequencies are transmitted with reference \textit{Magnitude} and \textit{Phase} of $|0dB|\angle 0\degree$. The test signals are represented in a polar coordinate system in Figure~\ref{fig:systematic}. On the other hand, the received signals which are attenuated and distorted can also be represented as polar coordinate $|M_{Rx}dB|\angle P_{Rx}\degree$. The high-level goal here is to characterize the communication channel based on a set of received signals of $RX$. In the realm of telecommunications, the problem at hand here is categorized in the well-known sub-field of \textit{Channel Characterizations} \cite{yin2016propagation}. Although similar measurement methodologies seek to characterize the wireless channels \cite{vcoja2017channel,peng2016channel,chaibi2015uwb}, we apply the same measuring methodology in the context of IC characterization to extract information. In other words, the channel characterization, in this case, yields information about the internals of the IC (specifically the contents of the registers). Hence, a systematic SCA could be accomplished by analyzing the set of $|M_{Rx}dB|\angle P_{Rx}\degree$ with respect to the target state of the DUT during the measurements.

\section{Physical-Level Analysis of Impedance Leakage}

To perform SCA on the reflected signals from the chip's die, it is vital to understand how the transmitted carrier signals are modulated based on the physical characteristics (and specifically the contents of the registers) on the chip.
As described earlier, the system's PDN (to and from the chip) can be considered a communication channel in high frequencies between a transmitter and a receiver. The VNA in our system is both the transmitter and the receiver on this channel.
Therefore, the channel model is analogous to a RADAR channel model~\cite{fens2008channel}.
In this case, the transmitted signal $x(t)$ from the VNA  at frequency $f_i$ can be written as follows:

\begin{equation}\label{eq:signal}
x(t) = A\sin(2\pi f_i t)
\end{equation}

where A is the amplitude of the transmitted signal.
Considering the PDN of the chip as the entity that applies the desired information to the carrier $x(t)$, the amplitude and phase of $x(t)$ are reshaped.
Thus, the modulated signal $\tilde{x}(t)$, received  at the VNA can be described~\cite{trees2013detection}:

\begin{equation}\label{eq:recieved}
\tilde{x}(t) = \alpha_i A\sin(2\pi f_i (t-2T) + \phi_i) + n(t)
\end{equation}

where $\alpha$ is the attenuation or fading factor, $\phi$ is the phase shift, $2T$ is the round-trip time of the signal, and $n(t)$ is a Gaussian noise added to the signal.
In our case study, we experimentally show that the characterization of the PDN, which contains a data-dependent leakage, could be observed in the amplitude, phase, and round-trip time variations. However, since our analyzes in this paper are confined within the frequency domain data, we do not consider the round trip time as a leakage parameter.

\subsection{Realization of Virtual Probes}\label{sec:vprobe}
As indicated in our approach, the transmit carrier signals $x(t)$ are injected on a wide range of frequencies $f_i$. 
As covered in Section~\ref{S_Z_Params_General}, well-studied back-scattering PDN profiling confirms that even in nanometer-scale characterization, different circuit elements (i.e., inductors and capacitors) reflect (and forward) different portions of input energy in accordance with carrier frequency~\cite{ichihashi2019simple} due to non-linearity in their frequency response~\cite{billings2013nonlinear}. 
This frequency-dependent behavior of the elements in the microscopic scale is affected by physical and intrinsic (such as dimensions) features, which determines the resonate frequency of the elements~\cite{quek2015characterization}. Hence, exploiting the same observation in our scenario, different input frequencies ($f_i$) yield capturing the response of different elements on the DUT. 
In accurate terms, the reflected signal at a particular frequency contains a dominant response of specific elements. 

\begin{figure}[t]
    \begin{subfigure}[b]{1\columnwidth}
    \includegraphics[width=\linewidth]{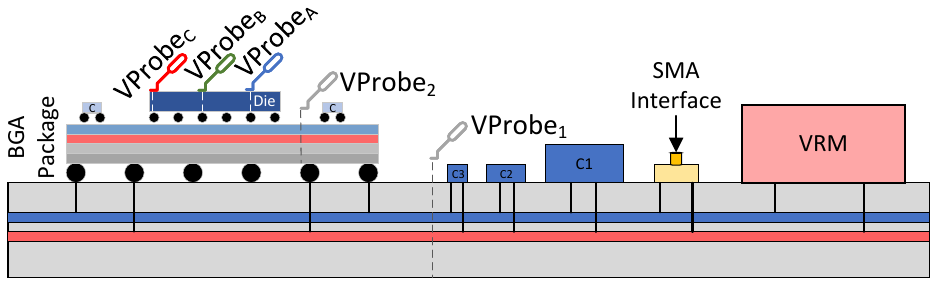}
    \caption{}
    \label{fig:vrm}
  \end{subfigure}
    \hfill 

  \begin{subfigure}[b]{1\columnwidth}
    \includegraphics[width=\linewidth]{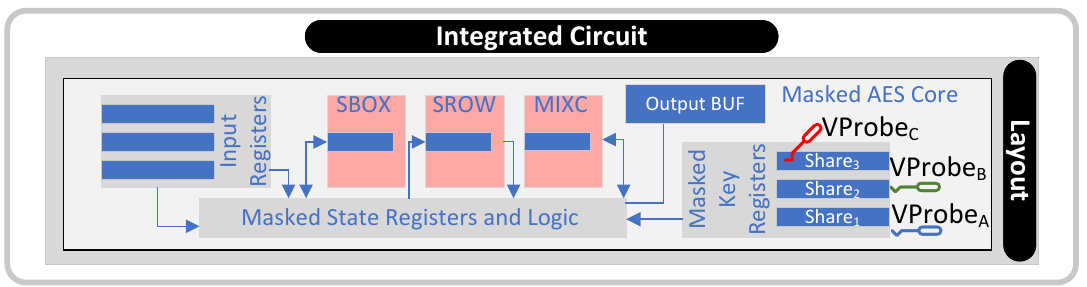}
    \caption{}
    \label{fig:layout}
  \end{subfigure}

    \caption{(a) Realization of virtual probes with the use of VNA and (b), Applying virtual probes on a masked AES implementation on a chip.}
    \label{fig:virtualprobe}
\end{figure}

To illustrate this, Figure~\ref{fig:vrm} shows a high-level cross-section block diagram of a simple PDN and a target IC. Here, an adversary generates $(x(t),f_i)$ using a VNA and injects the signals via the SMA interface. Based on the selected $f_i$ multiple \textit{Virtual Probes} could be deployed to characterize the board. For instance, on-board capacitors (e.g., $C1$ and $C2$) are large elements, and their dominant response could be observed in $\tilde{x}(t)$ with low-frequencies (\textit{MHz}). This indicates that any modification on these elements could be observed in  $\tilde{x}(t)$ with those frequencies.
In other words, by knowing those exact frequencies $f_{Probe_1}$, the adversary can place a \textit{Virtual Probe} on $C1$ and $C2$. Sweeping to higher frequencies (\textit{GHz}), the adversary will have multiple \textit{VProbes}, on different elements (physical locations) on the chip to perform a powerful SCA. 
This capability is granted by the fact that IC-level elements are physically asymmetric and placed and routed uniquely on the die.
Figure~\ref{fig:layout} depicts the utilization of \textit{VProbes} to effectively disable a \textit{Masked AES} implementation on a target chip. We show that, with a proper profiling process, an attacker can precisely determine frequencies for each share of the masked key to extract all secret values simultaneously. It is also noteworthy to mention that in Figure~\ref{fig:layout}, each target key register shows a dominant response on a specific frequency set, leading the adversary to distinguish the leakages.

\subsection{Effects of Parallel Computation and Masking Schemes}
Cryptography implementations in software fundamentally suffer from time-domain leakage caused by the serialized execution model. Many researchers have shown that even secured masking implementations could be easily broken by template attacks on software \cite{wu2022best,oswald2006template}.
In the case of masking implementations on hardware, with the assumption of parallel computing on shares, the data-dependent power leakage is experimentally shown to be reduced significantly~\cite{de2018hardware}. The CMOS power consumption model in masking schemes assumes that expected power consumption \textit{P} for a masked secret bit \textit{$c=(m,c\oplus m)=(s_1,s_2)$} hold the following condition \cite{de2017does}:
\begin{equation}\label{eq:mask_power}
\begin{split}
&P(c=0)=P(c=1)\\
&P(s_1=0,s_2=0)+P(s_1=1,s_2=1)=\\
&P(s_1=1,s_2=0)+P(s_1=0,s_2=1)
\end{split}
\end{equation}
Although decoupling effects~\cite{de2017does} and higher order attacks~\cite{standaert2005masking} can exploit the aforementioned condition, considering the CMOS \textit{Hamming Weight} model and as long as $s_1$ and $s_2$ computations are performed simultaneously at $t_1$ this constraint in Equation \ref{eq:mask_power} is believed to be maintained.
Compared to 1-d scalar power consumption measurements, impedance leakage is measured over set of $N$ samples of frequencies, making the leakage variable a 2-d parameter.

By applying the same condition on \textit{Impedance} leakage, an ideal one-bit masking rules the following: 
\begin{equation}\label{eq:mask_im}
\begin{split}
&\forall f_i \in F=\{BW\}:\mathcal{Z}_{f_i}(c=0)=\mathcal{Z}_{f_i}(c=1)\\
\end{split}
\end{equation}

As thoroughly investigated in Section~\ref{sec:vprobe}, constraints indicated in Equation~\ref{eq:mask_im} could not be trivially held by first order boolean masking, as the unique physical realizations, corresponding wiring and required routing for $s_1$ and $s_2$ computations yield in frequency-dependant measurements. Here in contrast to scalar power leakage model, impedance profile in Equation~\ref{eq:mask_im} should be satisfied for all frequency stamps to ensure the security. For instance, if $N=100$ impedance measurements are conducted from the bandwidth $F=BW=\{1GHz-2GHz\}$, Equation~\ref{eq:mask_im} should be held. In such strong constraint, one can derive $\exists {f_{\alpha}}\in F$, where $\mathcal{Z}_{f_\alpha}(c=0)\neq\mathcal{Z}_{f_\alpha}(c=1)$ and therefore impedance of masks values could be distinguished effectively.

\section{Proposed Attack Scenarios}

In this section, we will elaborate on multiple attack scenarios based on the developed scatter profile of the DUT. Ultimately, here we showcase that the proposed methodology could challenge \textit{t}-probe security 
model. Based on our findings we exhibit that unlike power consumption side channels, impedance leakage collected from scattering profiles does not scale down exponentially \cite{chari1999towards,ito2022success} by increasing the number of shares. 
On the contrary, we showcase that the impedance leakages of \textit{t-share} operands are leaked through distinguishable frequencies and do not cause significant additive noise to leakage measurement. 
In other words, although state-of-the-art masking schemes, such as Threshold Implementation~\cite{reparaz2015consolidating} and Domain-Oriented Masking~\cite{gross2016domain}, are highly effective in mitigating time-domain SCA (i.e., DPA on power consumption), they will not suffice against proposed frequency-domain impedance SCA. As thoroughly described in Section~\ref{sec:vprobe}, this is due to the fact that the physical characteristics of gate-level elements are scattered over a frequency bandwidth that can be captured by impedance profiling. Here, we exploit this fact to exhibit that the aforementioned characteristic is indeed data-dependent and can be used to reveal secrets.


We start off with simple case studies to verify scattering leakage is exploitable, and then we move toward realistic scenarios where sensitive data are protected by higher-order well-known masking schemes. 
First, we describe how adversaries can mount conventional non-profiled attacks via impedance analysis. Particularly we develop \textit{Differential Impedance Analysis} (DIMA) to break unprotected cryptographic implementations. 
Furthermore, a non-profiled \textit{Correlation attack} is also presented to illustrate conventional leakage models (i.e., Hamming Weight) are also effective in impedance analysis.

After verifying the applicability of naive non-profiling scenarios, we specifically aim to sidestep protected hardware implementations by exploiting on-die location-based profiling. 
As our main attack, we present \textit{Template Impedance Attack (TIMA)} to extract time-constant high-order masked operations. We showcase that \textit{TIMA} effectively breaks state-of-the-art masking schemes.

\subsection{Threat Model} DIMA and CIMA attacks are performed in known plaintext scenarios. For TIMA, we also assume the access to profile the random shares (i.e., all key shares). Building templates for mask registers plays a crucial role in mounting a successful template attack which is usually not taken into account. For a thorough discussion, please refer to~\cite{bronchain2021give}. 

On the execution level, since our method aims to extract data directly from the DUT, we assume that the measurements are performed when target data remain unchanged in some registers on DUT. 
This consideration makes our treat model very similar to static power side-channel analysis~\cite{moradi2014side,del2015side,moos2019static,moos2019staticb} and LLSI attack~\cite{krachenfels2021real,krachenfels2021automatic}, where instead of performing the measurements during the secret-dependant operations, adversary snapshots at timestamps where secret-dependant data are stored in some form (i.e., in Flip-Flops) on the DUT. In this regards, TIMA does not apply any time-series multi-variant analysis on the target like the works in \cite{del2015side,durvaux2016improved}. and the measurements are captured on a single clock-cycle. Here, we make use of multivariate frequency analysis which exploits the frequency-dependent leakage on the same time-stamp. Hence, our threat model requires a clock-controlled environment for high-speed DUT during the measurement. However, this limitation might be resolved by iterating the measurements without clock halting. (For more discussion please see Section~\ref{sec:clock})

\subsection{Naive Impedance Attacks}
In order to showcase the exploitability of impedance side-channel leakage, we demonstrate that conventional power-side channel attacks could be easily modified to be applied to impedance measurements. 
Here, we describe two high-level attacks useful for unprotected cryptographic algorithms. Namely, in the following, we describe impedance-based DPA and CPA.

\subsubsection{Differential Impedance Analysis}

Differential Power Analysis~\cite{kocher1999differential} on cryptography implementation exploits the variations of dynamic power consumption of DUT by employing hypothesis-based differential measurements of traces. 
For correct hypothetical secrets, the differential analysis maximizes a secret-dependant intermediate operation over time-domain measurement. 
We inspire the same methodology to deploy Differential Impedance Analysis (DIMA) over the frequency domain. In other words, for a correct hypothesis of a secret, the absolute difference of impedance measurements of a secret-related intermediate value should be maximized at some frequency stamp. This frequency stamp is physically related to the characteristics of the element that somehow(i.e., stores or transfers) intermediate value. The routine for DIMA is mostly similar to conventional DPA. 
The adversary could follow the same algorithmic process in DPA~\cite{kocher2011introduction}. However, instead of calculating differential measurement on time stamps, the attacker performs the differential analysis on frequency stamps. 
Algorithm~\ref{alg:1} illustrates a high-level description of DIMA. In Algorithm~\ref{alg:1}, \textit{$Trc_{a}[]$} represents the array of all measured traces for the attack and \textit{$Inp(Trc_{a}[i])$} shows the associated input with trace $Trc_{a}[i]$. Furthermore, \textit{$Int_{v}(key=k,input=i)$} depicts the result value of a selected intermediate bit value in the targeted cipher where key and input of the cipher are k and i, respectively.

\begin{algorithm}
\caption{Differential Impedance Analysis}\label{alg:1}
\begin{algorithmic}[]

\Function{DIMA Attack}{$Trc_{a}[]$} 

\For {$k\in K=\{0, 1, ..., 2^m\} $ }\Comment{$K$ is set of  key values}
\For {$i=0, 1, ..., |Trc_{a}|$}
\State $H \gets Int_{v}(k,Inp(Trc_{a}[i]))$\Comment{ intermediate value}
\If {$H=1$}
\State{$One_{Trc} \gets Append(Trc_{A}[i])$}
\Else 
\State{$Zero_{Trc} \gets Append(Trc_{A}[i])$}
\EndIf
\State\textbf{end if} 
\EndFor
\State\textbf{end for}   
\begin{tcolorbox}[colback=red!5!white,colframe=red!75!black]
\State $Diff_{k} \gets DOM_{freq}(Zero_{Trc},One_{Trc})$
\end{tcolorbox}

\EndFor
\State\textbf{end for} 

\State \textbf{return} $ArgSort(Mean(Diff,freq))$
\EndFunction

\end{algorithmic}
\end{algorithm}

As indicated in Algorithm~\ref{alg:1}, the Difference Of Mean ($DOM_{freq}$) is performed element-wise on each frequency stamp of impedance traces. Finally, we apply a simple averaging on each candidate to represent each guess key with a single score. Then an \textit{Argument Sort} on frequency-normalized $Diff_{k}$ values are performed and target key is extracted.
On the algorithmic level, the frequency indices of maximum values of $Diff_{Key=k}$ indicates the frequencies at which the intermediate value's physical deployment on the die leaks the most. 
Particularly, the same frequencies could be considered independently and be used as a profile to reveal this specific intermediate value's content. 
We take advantage of similar behavior to put together a powerful profiling attack which will be elaborated later on.

\subsubsection{Correlation Impedance Attacks}
As another well-established power side-channel attack, Correlation Power Analysis (CPA)~\cite{brier2004correlation} is also used, and they mitigate some drawbacks of DPA~\cite{le2006proposition}. CPA uses a power consumption model on the hardware to establish a correlation between the secret-dependent operation and inputs over power measurement. For instance, a famous metric assumes that power consumption of operands on the circuits tracks a linear trend with respect to its Hamming Weight (HW). Here, the same approach is utilized for the adversary to attack an unknown secret. We use frequency-stamped impedance traces to mount a Correlation Impedance Attack (CIMA). Although different leakage models for impedance traces could be employed as a proper candidate, we use a simple HW model to develop our attack. In CIMA attacker executes Pearson Correlation~\cite{cohen2009pearson} over frequency-domain impedance values. Hence, target intermediate content leaks on a specific frequency stamp that is indicated by CIMA. 
Armed with an HW model, we launch CIMA on the first round of AES's S-Box output to showcase a successful correlation impedance attack. It is noteworthy to mention that based on our evaluations (in Section~\ref{sec:CIMA}), the HW model shows a near-linear leakage trend over impedance's phase ($\angle S_{11}$) on the target frequency stamp.

Although CIMA successfully breaks the hardware realization of a prominent cryptographic algorithm (i.e., AES) with a much less number of measurements, it is not effective against masking schemes. This is due to the fact that CIMA like other correlation attacks, strives to discover the likelihood of a single secret-dependant intermediate value, which is largely diminished by masking~\cite{de2018hardware}. 
In the following, we propose an attack that builds upon frequency-spanned leakage of impedance measurements enabling us to extract masked secrets through frequency analysis in hardware implementations.    

\subsection{Template Impedance Attacks}
Template attack proposed by Chari et.al. \cite{chari2003template}, provides a strong attack methodology where the adversary can profile a target hardware of her choosing with an arbitrary cryptographic implementation (e.g., protected) and break similar hardware using a limited number of power consumption measurements. In contrast to non-profiling methods which strive to eliminate noise by averaging over large measurement traces, template attack utilizes multi-variant characterization of points of interest (including the noise) by employing the identical target hardware, making it extremely powerful during attack phase~\cite{chari2003template}. 
Owing to the nature of the impedance measurements, we believe a similar Template Impedance Attack (TIMA) is the most powerful attack that can be performed using impedance analysis.  
TIMA follows the same algorithm as a regular template attack while performing on frequency stamps. Algorithm \ref{alg:3} depicts the high-level flow of TIMA.

\begin{algorithm}
\caption{Template Impedance Attack}\label{alg:3}
\begin{algorithmic}[]

\Function{TIMA Profile}{.} 

\For {$k\in K=\{0, 1, ..., 2^m\} $ }\Comment{$K$ is set of  key values}
\For {$i=0, 1, ..., N_{p}$}
\State $Trc_{p}[k,i] \gets Measure(\{Z_{fr_0},Z_{fr_1},...,Z_{fr_f}\})$
\EndFor
\State\textbf{end for}   
\State $AVTrc_{p}[k] \gets Mean(Trc_{p}[k],N_{p})$
\EndFor
\State\textbf{end for}   
\begin{tcolorbox}[colback=red!5!white,colframe=red!75!black]
\State $DM \gets DOM(AVTrc_{p}[k],freq)$ 
\State $POI[p] \gets TopK(DM,p)$
\end{tcolorbox}

\For {$k\in K=\{0, 1, ..., 2^m\} $ }
\State $Mean_{p}[k] \gets Mean(Trc_{p}[k],POI)$
\State $Cov_{p}[k] \gets Cov(Trc_{p}[k],POI)$

\EndFor
\State\textbf{end for}   
\State \textbf{return} $Mean(Trc_{p}),Cov(Trc_{p})$

\EndFunction

\Function{TIMA Attack}{$Trc_{a}[],Mean_{p}[],Cov_{p}[]$} 
\For {$k \in H=\{0, 1, ..., 2^m\}$}\Comment{$H$ is set of guess target values}
\State $Dis_k \gets MultivarGaussian(Mean_{p}[k],Cov_{p}[k])$
\EndFor

\State\textbf{end for} 
\For {$i=0, 1, ..., N_{p}$}
\begin{tcolorbox}[colback=red!5!white,colframe=red!75!black]
\State $STrc_{a}[] \gets Select(Trc_{a}[],POIS)$
\end{tcolorbox}

\For {$k \in H=\{0, 1, ..., 2^m\}$}
\State $P[k] \gets PDF_{eval}(Dis_k,STrc_{a}[j])$
\EndFor
\State\textbf{end for} 
\State $Res[k] \gets Acc(P[k])$
\EndFor
\State\textbf{end for} 

\State \textbf{return} $ArgMax(Res[])$

\EndFunction
\end{algorithmic}
\end{algorithm}
 In Algorithm~\ref{alg:3}, each profile trace 
$Trc_{p}[key=k,Inx=i]$, is a collection of measurements associated with the impedance of the target at frequencies $\{fr_0, fr_1, ..., fr_f\}$, where the $k$ and $i$ refers to the cipher's key and trace index, respectively. Moreover, $Mean(.)$ and $Cov(.)$ are \textit{Average} and \textit{Covariance} calculation functions.
As demonstrated, the \textsc{TIMA Profile} phase is performed on the DUT controlled by the adversary. 
As highlighted, Points Of Interest (POIs) in this attack are determined by averaging measurement values over a frequency band. Hence, frequency stamps that cause maximum difference for all the possible target $\mathcal{K}$ are selected using the Difference in Means ($DOM$) metric. Moreover, we apply \textit{TopK}, a localized Top-K \cite{arai2007anytime} with empirically set smoothing factor $\alpha$ to select final POIs. Note that as collected scattering measurements are complex numbers, \textsc{TIMA Profile} could be applied to \textit{$\angle S_{11}$} or \textit{$|S_{11}|$} or even \textit{$f(|S_{11}|,\angle S_{11})$}. As discussed earlier in this article, we mainly focus on \textit{$\angle S_{11}$} profiling since it is more resilient to additive noise compared to \textit{$|S_{11}|$}. 
Nevertheless, one could deploy an iterative estimation \cite{rossi2018mathematical} or optimization \cite {bottou1998online} procedure  to find near-optimal parameters for \textit{$f(|S_{11}|,\angle S_{11})$} leakage profiling kernel to enhance the attack success rate.

In the second phase, the attacker has limited access to the target DUT and captures a small number of measurement traces of impedance \textit{$Trc_{a}[]$}. Consequently, \textsc{TIMA Attack} routine is executed. As highlighted, measurements on pre-determined frequency stamps (POIs from the profiling phase) are selected first ($STrc_{a}[]$). Then probability evaluation ($PDF_{eval}$) of each measured attack trace is computed for each key hypothesis based on its Gaussian multi-variant distribution (\textit{$Dis_k$}) extracted during \textsc{TIMA Profile} phase. Lastly, probabilities are accumulated for all captured traces to indicate the final candidate. 

As will be explored, TIMA can be successfully applied in multiple attack scenarios and particularly on masked implementation. Specifically, we exercise a single bit TIMA on masked AES implementation (in Section~\ref{tima_aes}) to recover all masked shares of the key, bit by bit. On the algorithmic level, the advantage of TIMA comes from the fact that every single bit on DUT contributes to a unique set of \textit{POIs} over the frequency band, and consequently, forms a (fairly) distinguishable Gaussian multi-variant distribution that can be estimated with a fair amount of profiling measurements.

\section{Evaluation}\label{sec:Experimental_setup}
\subsection{Experimental Setup}

\begin{figure}[t]
  \begin{subfigure}[b]{0.54\columnwidth}
    \includegraphics[width=\linewidth]{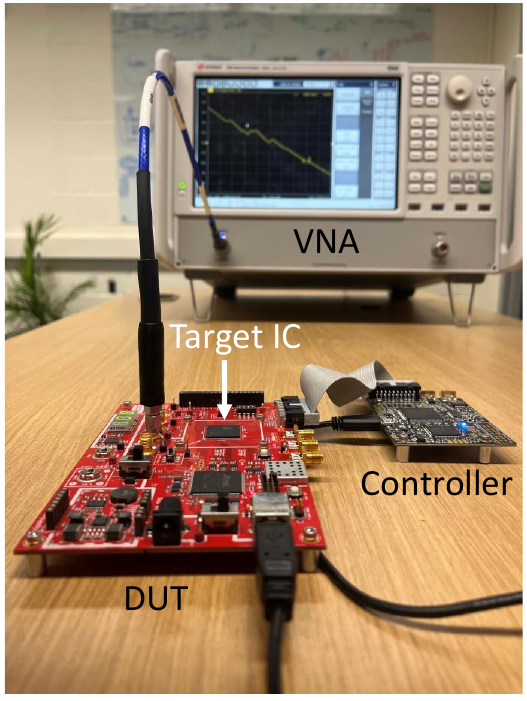}
    \caption{}
    \label{fig:experiment_set_photo}
  \end{subfigure}
  \hfill 
  \begin{subfigure}[b]{0.45\columnwidth}
    \includegraphics[width=\linewidth]{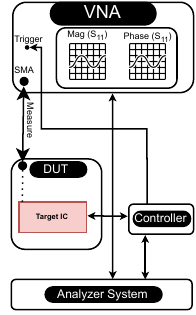}
    \caption{}
   \label{experiment_set_diagram}
  \end{subfigure}
    \caption{Measurement setup. (a) VNA capturing \textit{$S_{11}$} traces from the DUT  and (b) Experimental setup diagram.}
  \label{fig:experiment_set}
\end{figure}


\subsubsection{Measurement Equipment}\label{subsec:Measrement_setup}
We utilized a Keysight ENA Network Analyzer E5080A\cite{enaKeysight}, which enables RF/microwave scattering measurements and operates on 9KHz - 6 GHz frequency bandwidth.
We used Minicircuit CBL-2FT-SMNM+ characterization shielded cables \cite{minicircuits} suitable for scattering measurements which are also operable in the same frequency bandwidth.
The used VNA ports have internal capacitors to filter out the DC voltage on the $V_{CCINT}$, and therefore, no Bias Tee is needed.

\subsubsection{Device Under Test}\label{subsec:DUT}
For our experiments, we used NewAE CW305 board (NAE-CW305)~\cite{CW305}, which is equipped with an AMD/Xilinx Artix-7 FPGA (XC7A100T), built with a 28 nm technology, see Figure \ref{fig:experiment_set_photo}.
CW305 board provides direct access to the FPGA's PDN network, which was the main reason for the selection of this board.
Moreover, while the FPGA contains multiple PDN domains (e.g., $V_{CCINT}$, $V_{CCO}$,  ) a 1V domain supplying the core ($V_{CCINT}$, $V_{CCAUX}$, etc.), for our evaluations, $V_{CCINT}$ power domain is our primary target PDN as it is connected the FPGA registers.
Furthermore, CW305 has multiple SMA (SubMiniature version A) connectors that enable access to a shunt resistor, as well as a 20 dB low-noise amplified low-side signal suitable for power analysis. 
However, our experiments are carried out by the SMA port on the low side of the shunt resistor {X3 port on the board}, which gives us direct access to the PDN of the FPGA.

\subsubsection{Analyzer and Controller Configuration} 

To control the state of the target FPGA chip, we have utilized a NewAE CW-Lite board~\cite{cwlite}, which provides serial communication with the DUT and could be used as an intermediate controller to transfer plaintext and receive ciphertext from the target IC. Furthermore, to conduct measurements, the CW305 board is configured to synchronize IC's clock once the controller receives the trigger signal (e.g., CW-Lite). Specifically, for our clock controlled experiments (e.g. TIMA), target's clock signal is generated via PLLs on CW305 and a feedback is sent to the controller at the same time. Based on the received and synchronous clock signal of the target and upon reaching to the desired time-stamp, a mask signal is sent by the controller and target's clock signal is masked and computation is halted. During this idle status, although the PLL clock on the board is oscillating but the target clock on the IC is gated. Consequently, a trigger signal is sent to the VNA for the measurement during this period. In our experiments, PLL board clock is set to $100 MHz$.

In order to schedule and prepare the input data for the IC and also analyze measurements and conduct our attacks, we employed \textit{Python 3.7} scripts. 
We used \textit{PyVisa} and \textit{Scipy.Stat} to communicate with the instruments and perform statistical analysis, respectively. 
Furthermore, hardware designs are written in Verilog, and synthesizes are carried out by Xilinx Vivado. 
The \textit{Analyzer System} is a machine with an Intel XEON E5 2697 V3 CPU clocked at 2.6 GHz, equipped with 128 GB
of DDR3 RAM, and runs an Ubuntu 20.04.6 LTS.  

\subsubsection{Measurement Procedure}
Figure~\ref{experiment_set_diagram} depicts our experiment diagram. Our experiments process could be described as follows:

\begin{enumerate}
  \item[$\blacksquare$] As the initial step, The desired hardware design of the \textit{Target IC} (e.g., a masked AES) as a bitstream is programmed using a JTAG connection.
  \item Arbitrary input data (i.e., masked plaintext, masked keys, etc.) are prepared in the \textit{Analyzer System} and are sent to the \textit{Controller}.
  \item Using a serial interface \textit{Controller}, sends the data to the \textit{Target IC} and collects timing stamps (Clock triggers) from \textit{Target IC} during its operation. At a desired time stamp, it triggers the VNA for the measurement.

  \item VNA performs a measurement upon getting triggered and sends back the measured traces to the \textit{Analyzer System}.
  \item Once the execution on the \textit{Target IC} is finished, output is received by the controller and is sent to \textit{Analyzer System} via UART connection for verification purposes.

\end{enumerate}

For the suitable profiling procedure required for our proposed attacks, we developed an iterative and automated process to capture the traces. Furthermore, note that in accordance with our threat model, our attacks require time-constant measurements as it exploits the physical characteristics of the die. 
Hence, synchronization and clock control are vital in our measurements, handled via \textit{Controller} in our attack scenarios.

\subsection{Target Implementation and Configuration }
{
\textbf{VNA Configurations and Frequency bands.} On each set of experiments, based on the target implementation different frequency target band is selected. Based on our experiments, different implementations on the FPGA results in different leakage on each narrow-band frequency. This is due to the variations in physical realization of components on the target which effects the impedance profile. We indicate that different target frequency band might result in different (maybe superior) attack success rate. Our targeted bands are selected naively and experimentally. Furthermore, \textit{IF Bandwidth} and \textit{Averaging} factors, used for adjusting the VNA measurement quality are set experimentally in our attacks. IF Bandwidth is set to \textit{500 Hz} in our experiments to reliably filter out unwanted responses such as higher frequency spectral noise \cite{keysightman} and Averaging factor is selected to be $AV_{idx}=200$ in TIMA attack to reduce measurement noise floor. \\
\textbf{Implementation of AES S-box.} Unprotected AES targeted for CIMA and DIMA in our experiments is implemented with a fully pipelined architecture using Canright’s S-box \cite{canright2005very}. These attacks are carried out at the first clock-cycle (first round S-box's output).} \\
{
\textbf{Implementation of Masked AES.} Ultimately we deploy experiments with TIMA, targeting AES DOM implementation for FPGA \cite{DOM2016} with 3 shares (masking order of 2). We mount two attacks on different scenarios. 1) The measurements are performed on the first round's key-share byte registers of extended Canright’s masked S-box\cite{canright2008very} which is the underlying S-box circuitry used in AES DOM. This attack targets the first byte of the key and is carried out when three key-share registers are loaded in the core before the execution of the first masked S-box (clock-cycle 1)  2) We target the full key extraction during the initialization process and once key values ($3 \times 128$ bits) are loaded into the registers. We deploy TIMA before starting AES-DOM is executed (Where $aes\_start$ signal is set at clock-cycle 0).\\
}
{
\textbf{Complex Number Analysis.} Although the impedance values (alternatively $S_{11}$) are complex values which are usually represented by magnitude and phase numbers, our main attacks (i.e., CIMA, DIMA, and TIMA) are only carried out on the phase (\textit{$\angle S_{11}$}) part of the measurements.
}

\subsection{Attacking Fan-out Registers}
We start off with a simple profiling analysis of a design with fan-out registers. 
The same methodology is used by Moradi et al.~\cite{moradi2014side} and similar research scenario~\cite{moos2019staticb} to analyze the leakage of static power consumption. Here, we cascaded two 1024 sets of FDCEs ( D Flip-Flop with Clock Enable and Asynchronous Clear). In contrast with static-based power SCA~\cite{moradi2014side}, we explicitly specified the location~\cite{consXilinx} of the register sets for the implementation phase as different locations of registers yield different results in our evaluation. In this experiment, the wiring and locations are manually specified to deploy two connected register sets on two adjacent FPGA slices.  

\textbf{High Fan-out Binary Registers}\label{Sec:fan-out}
In the first experiments, we program the DUT to set all 2048 registers to either \textbf{\texttt{0b0}} or \textbf{\texttt{0b1}}. 
This setting is done via deploying a control register on the FPGA configured by the \textit{Controller}. 
Upon receiving the \textit{run} command from the \textit{Controller}, values on the registers are set. 
For each case, we collect the total number of \textit{600} traces in the frequency range of \textit{$F=2.7 GHz-3 GHz$}, with 5000 linearly spanned frequency stamps. In other words, $((3-2.7)/1000)\times10^9 Hz=300 kHz$ is set as the frequency resolution.
To minimize temperature-induced drifts, we follow a normalization process where we perform a  reference measurement (e.g. a FPGA register fan-out program where all registers are cleared.) after each measurement and store the difference as the final trace. Note that since our attack is a differential attack, normalized values do not affect the final outcome. 

\begin{figure}[t]
  \begin{subfigure}[b]{0.49\columnwidth}
    \includegraphics[width=\linewidth]{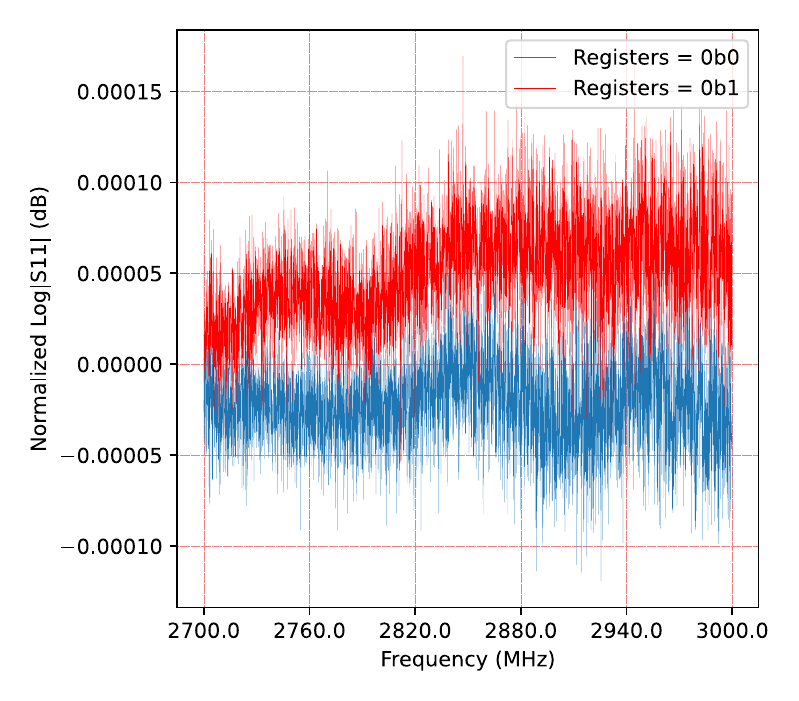}
    \caption{}
    \label{fig:reg_mag_abs}
  \end{subfigure}
  \hfill 
  \begin{subfigure}[b]{0.49\columnwidth}
    \includegraphics[width=\linewidth]{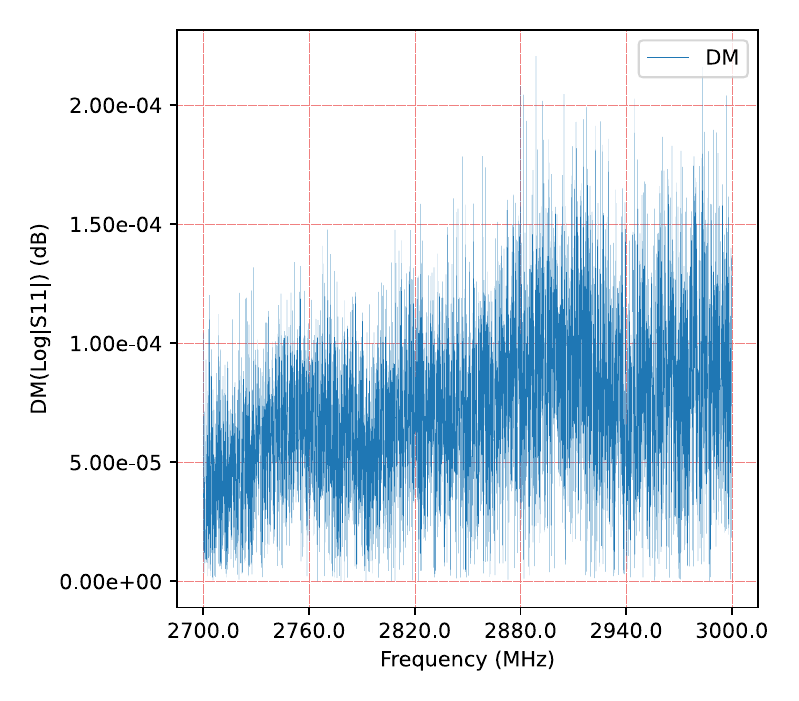}
    \caption{}
   \label{fig:reg_mag_dm}
  \end{subfigure}
    \caption{$|S_{11}|$ leakage on fan-out register. (a) Normalized and averaged comparison (b) Differential analysis of leakage.}
  \label{fig:reg_mag}
\end{figure}

Figure~\ref{fig:reg_mag_abs} depicts average values of \textit{$S_{11}$} magnitude for each case of the experiment over the selected frequency band. 
The difference (\textit{DM}) between the two groups is illustrated in  Figure~\ref{fig:reg_mag_dm}. 
As shown in this figure, the difference exists over the entire selected frequency band, however, at some stamps, it is larger compared to others. 
This indicates that at certain frequency stamps, the magnitude of the impedance differs based on the content of the registers and their corresponding wiring on specific positions on the die.

\begin{figure}[t]
  \begin{subfigure}[b]{0.49\columnwidth}
    \includegraphics[width=\linewidth]{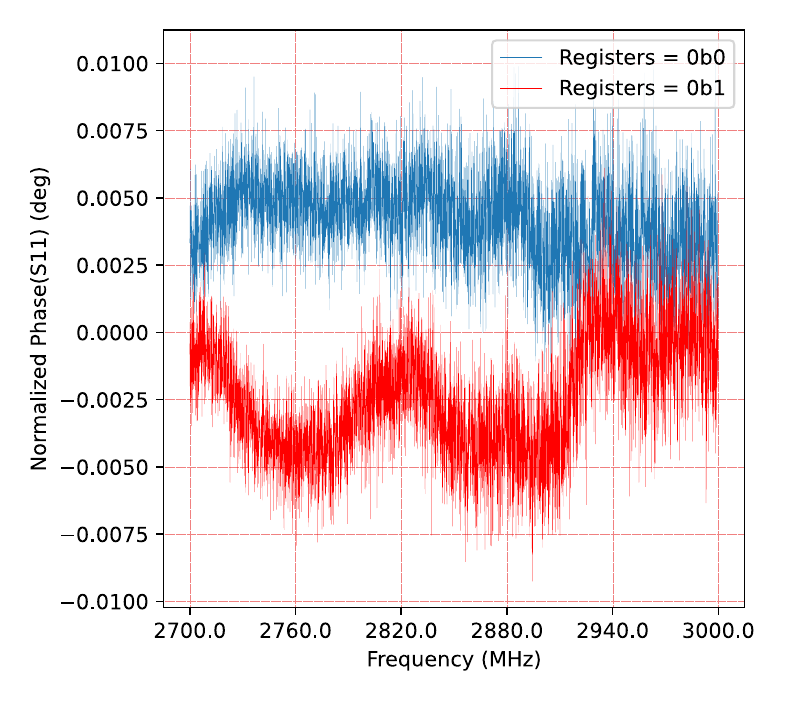}
    \caption{}
    \label{fig:reg_phas_abs}
  \end{subfigure}
  \hfill 
  \begin{subfigure}[b]{0.49\columnwidth}
    \includegraphics[width=\linewidth]{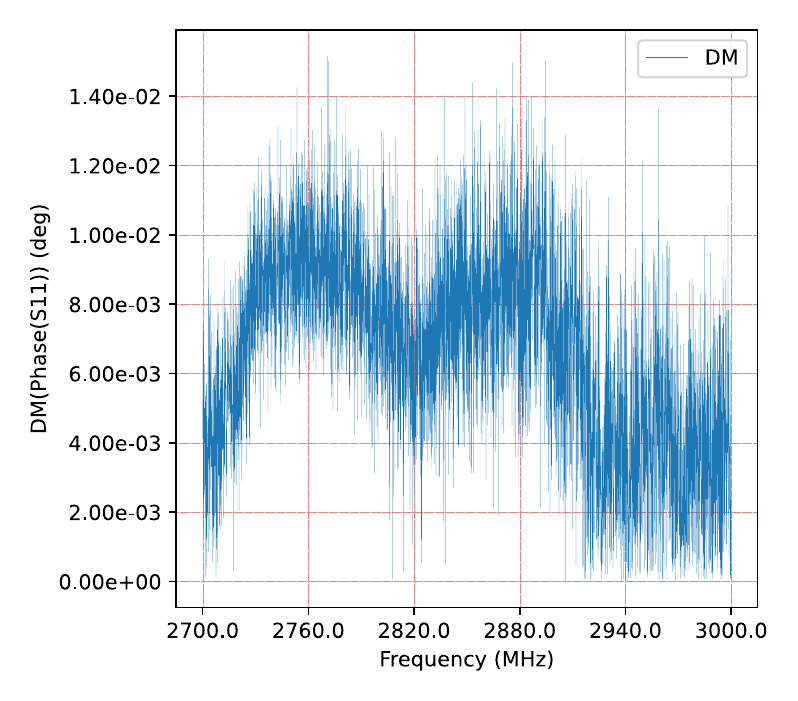}
    \caption{}
    \label{fig:reg_phas_dm}
  \end{subfigure}
    \caption{$\angle S_{11}$ leakage on fan-out register. (a) Normalized and averaged comparison (b) Differential analysis of leakage.}
    \label{fig:reg_phas}
\end{figure}

Similarly, Figure~\ref{fig:reg_phas} details the average difference of \textit{$\angle S_{11}$} for each group of traces. 
Note that \textit{$\angle S_{11}$} and \textit{$|S_{11}|$} are uniquely different metrics with completely different behavior. Consequently, each case in Figure~\ref{fig:reg_mag_abs} and Figure~\ref{fig:reg_phas_abs} follow different pattern. For instance, normalized \textit{$\angle S_{11}$} for case \textbf{\texttt{0b1}} is less than \textbf{\texttt{0b0}} traces for most of the selected frequency band, where for \textit{$|S_{11}|$} opposite behavior is observed. Also, the absolute difference at some frequencies is larger compared to the magnitude DM. This can be justified as often RF \textit{Phase} measurements are known to be more resistant to noise compared to magnitude.



\subsection{Attacking AES S-BOX}
\begin{figure}[t]
   \includegraphics[width=0.7\linewidth]{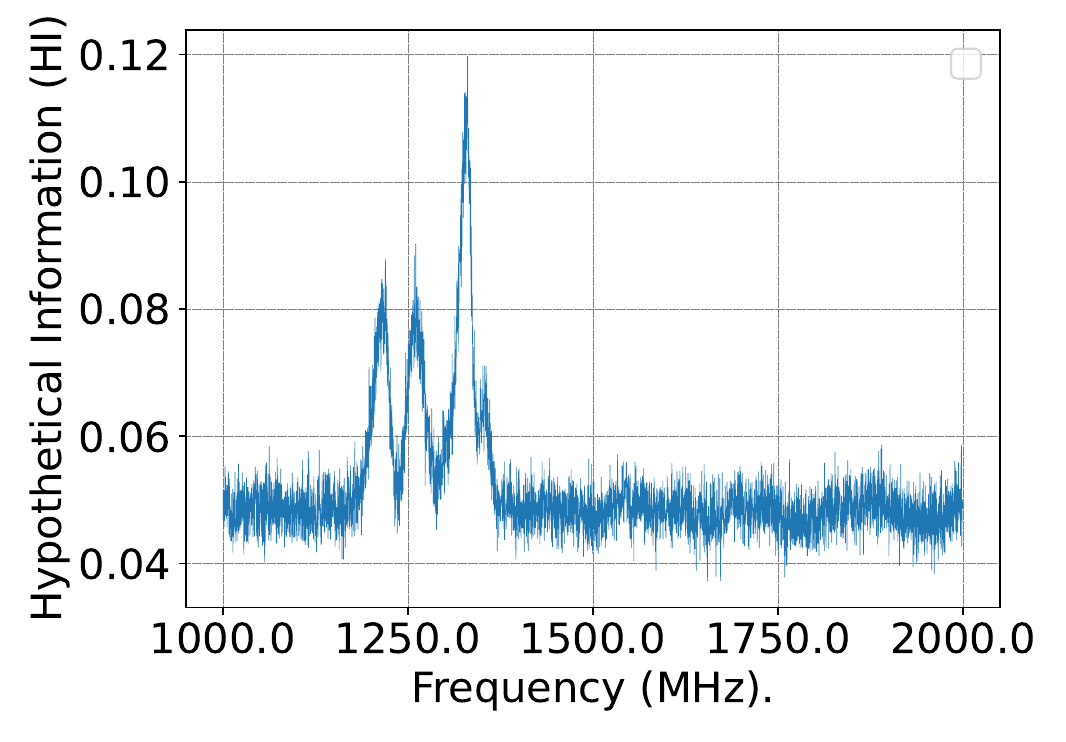}
	\caption{Hypothetical Information (HI) the impedance leakage for the presented S-box attack.}
            \label{fig:hi}                 
 \end{figure}

As the next step, we advance our target to attack AES S-BOX. 
We consider first-round S-BOX with a known plain-text scenario. 
In the following, we apply CIMA and DIMA attacks to an unprotected AES S-BOX implementation (derived from ProjectVault~\cite{vault}). Also, note that S-BOX's output $Tar=S-BOX(K_{in},P_{in})$ is considered as the target intermediate value in our attacks, and the goal is to find the first byte of the secret key value.  
{In order to verify the leakage for our captured traces, we first demonstrate an information leakage analysis. Figure \ref{fig:hi} shows the Hypothetical Information (HI) \cite{bronchain2019leakage} of impedance leakage for the S-box target in frequency domain. It shows that the some frequency sub-bands are more exploitable compared to others.
}
 
\begin{figure}[h!]
  \begin{subfigure}[b]{0.50\columnwidth}
    \includegraphics[width=\linewidth]{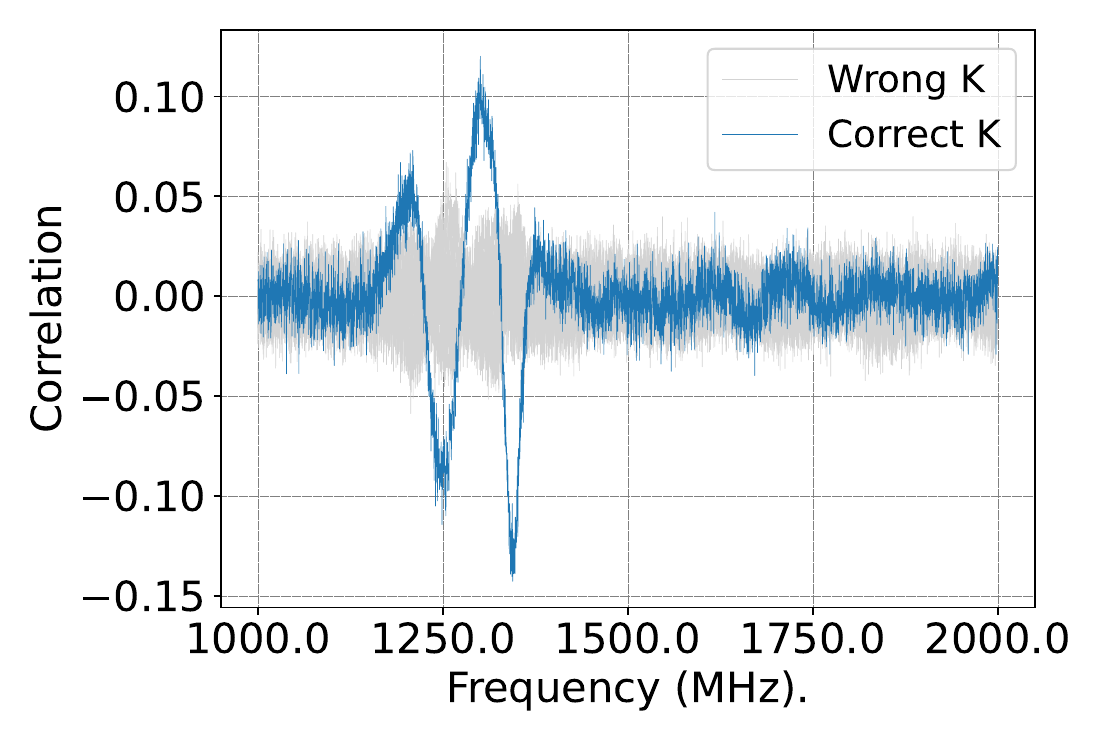}
    \caption{}
    \label{fig:mag_reg_att}
  \end{subfigure}
  \hfill 
  \begin{subfigure}[b]{0.49\columnwidth}
    \includegraphics[width=\linewidth]{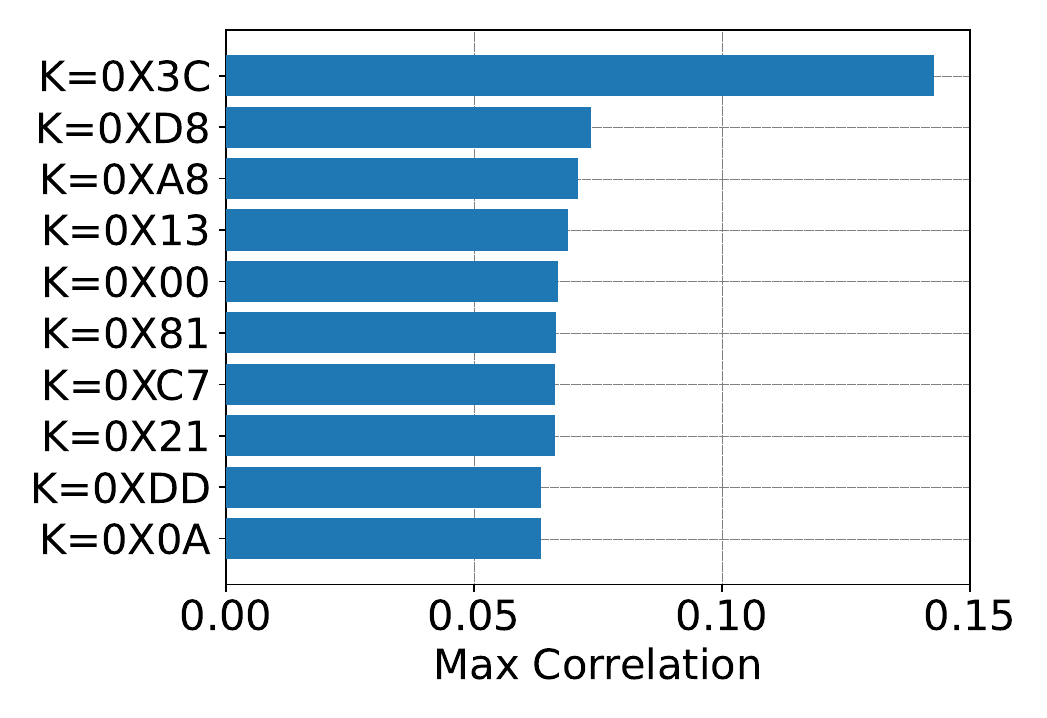}
    \caption{}
    \label{fig:pha_reg_att}
  \end{subfigure}
    \caption{\textit{CIMA} on 1200 samples (a) Correlation index over selected frequency band  and (b) Top correlated keys and the maximum value of correlation (the selected frequency for keys are not necessarily the same).}
    \label{fig:cpa}
\end{figure}

\subsubsection{CIMA}\label{sec:CIMA}
As our first attempt to break AES using impedance data, we deploy \textit{CIMA} based on a conventional HW model. In this experiment, we mount our attack on 1200 measurement traces. Furthermore, we select the frequency range of \textit{$F=2GHz-3GHz$}, with 3000 linearly spanned stamps. Figure~\ref{fig:cpa} depicts the results of CIMA in terms of correlation index.

\begin{figure}[t!]
  \centering \noindent
   \includegraphics[width=.9\linewidth]{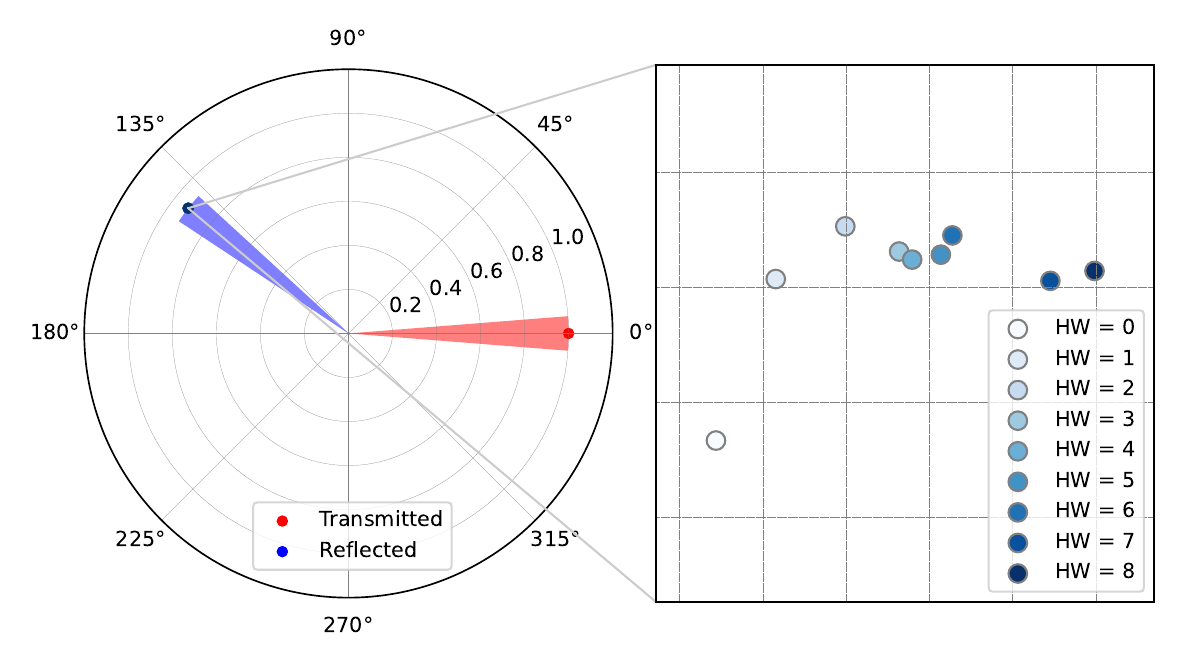}
	\caption{Polar representation of each \textit{HW} group of the reflected traces at $Freq=1328 MHz$, where CIMA score is maximized. Near-linear trend of HW leakage is visible for $\angle S_{11}$. }
            \label{fig:polar}                 
 \end{figure}

Furthermore, to verify that the chosen model in our attack successfully distinguishes on \textit{Magnitude} and \textit{Phase}, Figure~\ref{fig:polar} represents the polar distribution of the collected measurements with respect to the \textit{HW} of the chosen intermediate value. As shown, we grouped up and averaged the traces in their corresponding \textit{HW} class. 
The distribution of reflected signals (indicated in blue) shows that groups could be effectively distinguished.  

\subsubsection{DIMA}
\begin{figure}[h!]
  \begin{subfigure}[b]{0.49\columnwidth}
    \includegraphics[width=\linewidth]{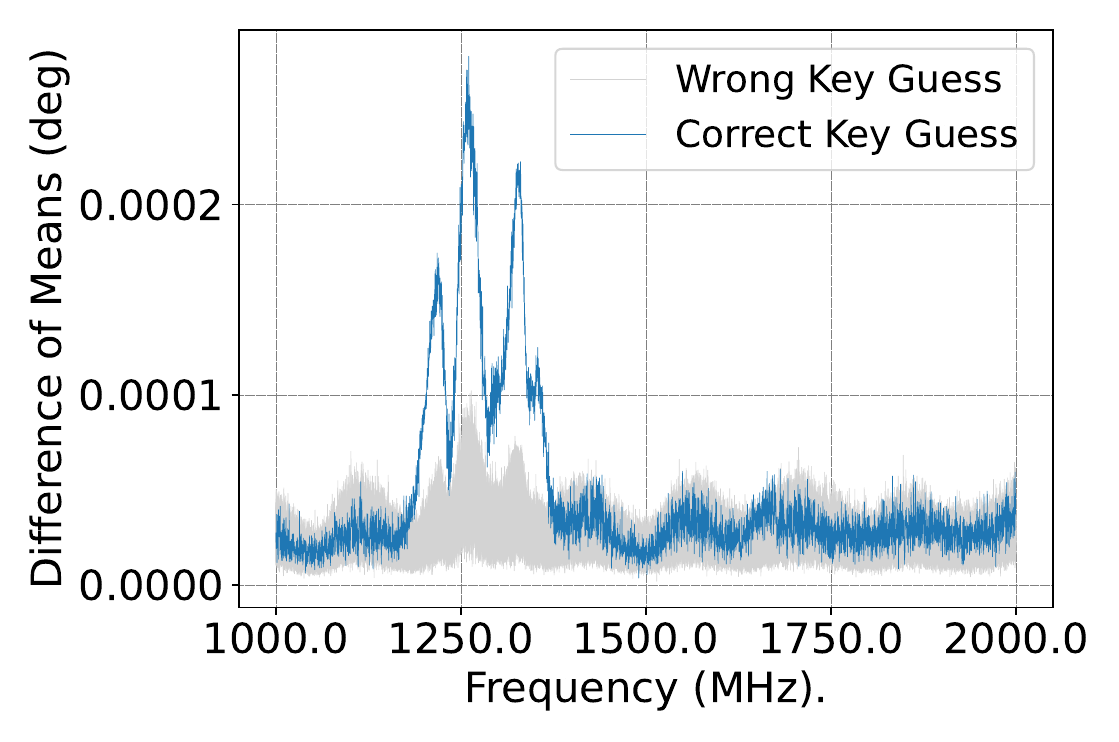}
    \caption{}
    \label{fig:mag_reg_att}
  \end{subfigure}
  \hfill 
  \begin{subfigure}[b]{0.50\columnwidth}
    \includegraphics[width=\linewidth]{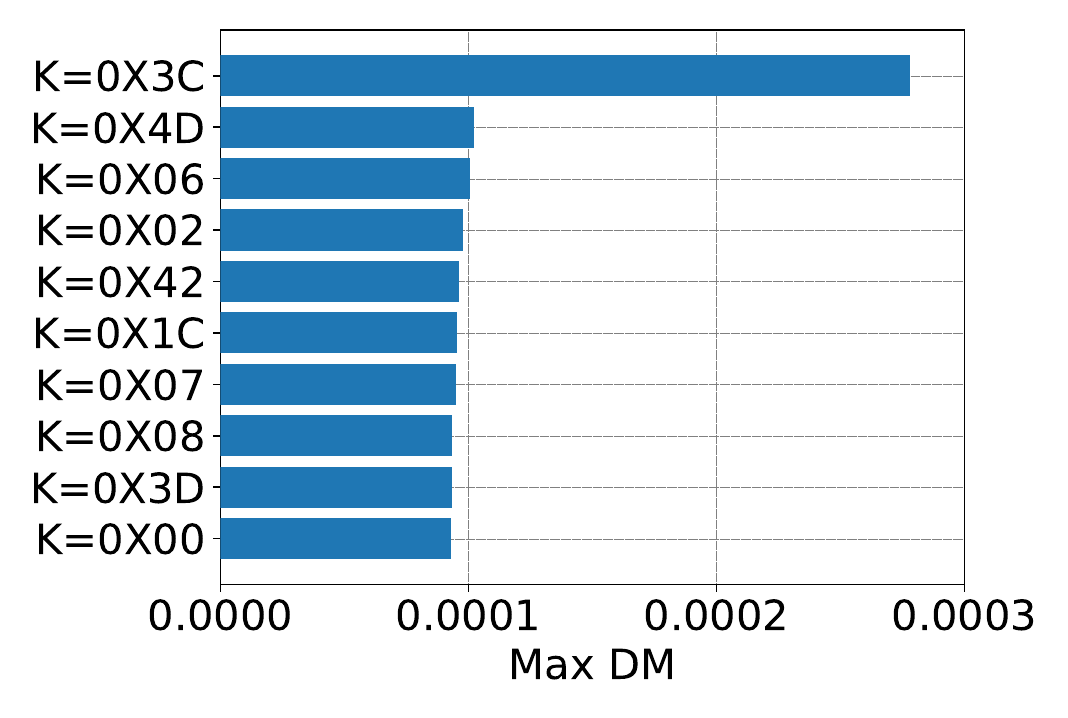}
    \caption{}
    \label{fig:pha_reg_att}
  \end{subfigure}
    \caption{\textit{DIMA} on 3000 samples (a) Differential results over the key space (b) Keys with the maximum value of multi-bit differences.}
    \label{fig:DIMA}
\end{figure}

Our next attack scenario is to perform impedance differential analysis (namely \textit{DIMA}) on an AES S-Box. 
For this attack, we employ $N=3000$ traces on  the frequency band of \textit{$F=1 GHz-2 GHz$}, with 3000 frequency samples at each measurement. Figure \ref{fig:DIMA} shows the final differential results for the possible key space using a multi-bit DIMA analysis.

\begin{figure}[!h]
  \begin{subfigure}[b]{0.49\columnwidth}
    \includegraphics[width=\linewidth]{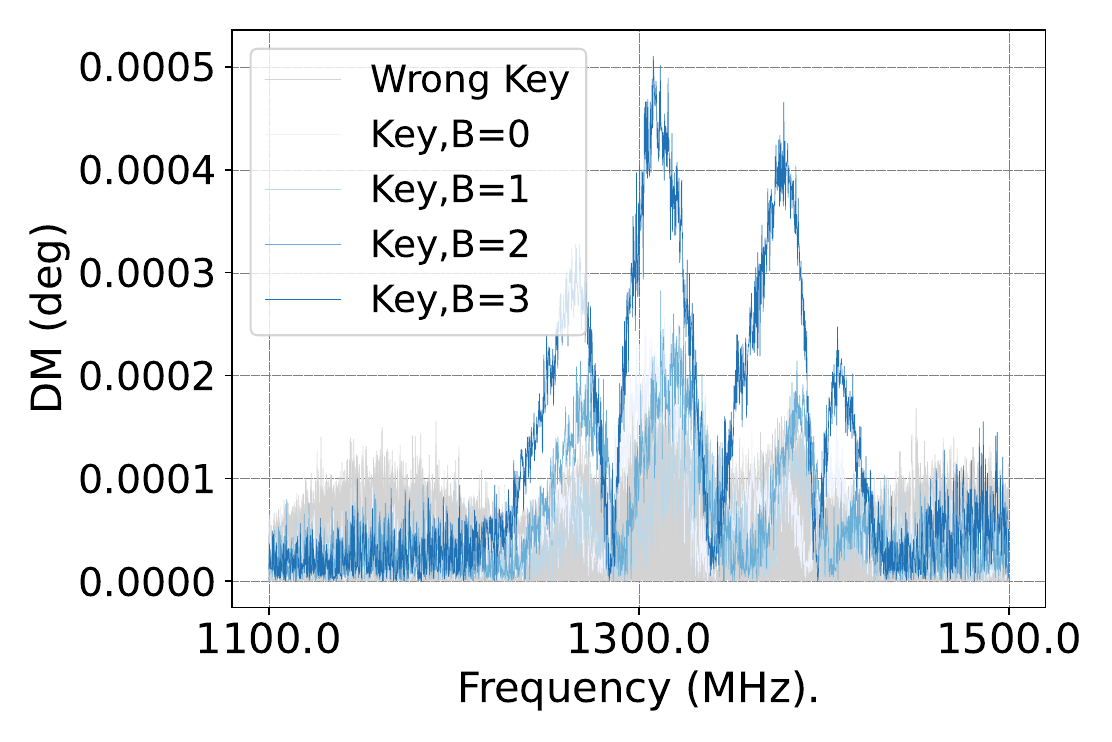}

  \end{subfigure}
  \begin{subfigure}[b]{0.49\columnwidth}
    \includegraphics[width=\linewidth]{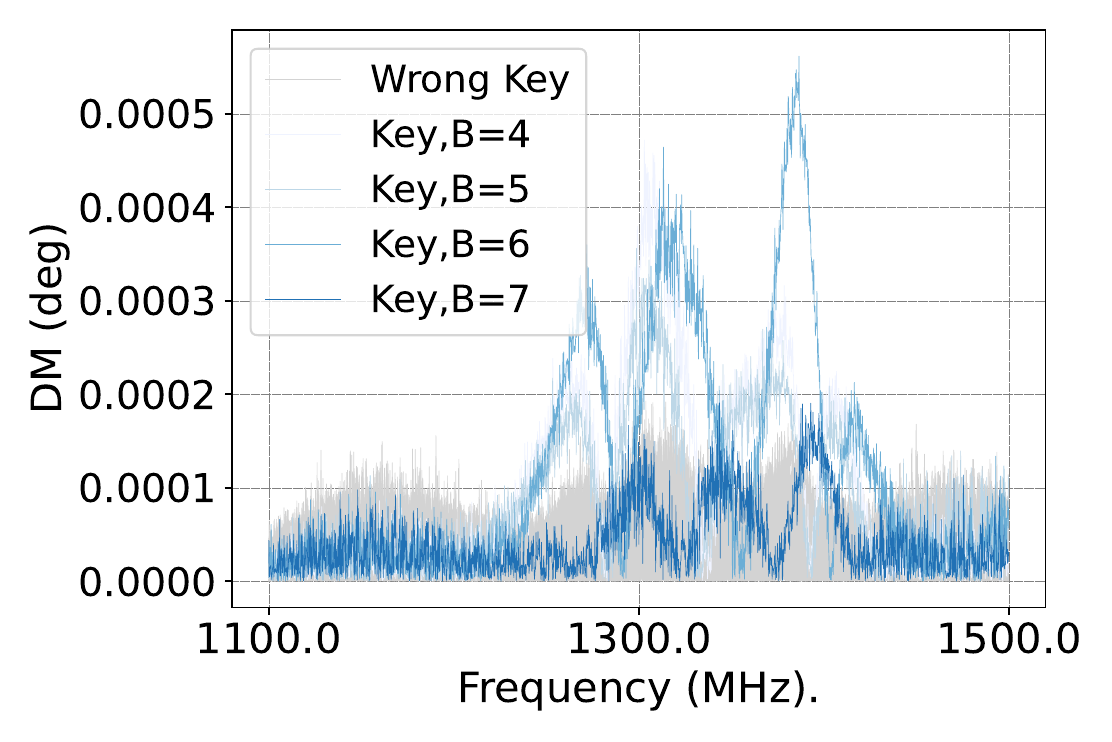}

  \end{subfigure}
    \caption{DIMA leakage analysis of different bits of the intermediate value}
    \label{fig:dima_bit}
\end{figure}

As another analysis, we investigated and measured the leakage of each individual intermediate bit as our indicator. Figure~\ref{fig:dima_bit} depicts the leakage for different bits of the target intermediate value. Although some bits are potentially more exploitable for the attack, it is clearly observed that each individual bit leaks in a distinguishable set of frequency samples.

\subsection{Attacking DOM AES}\label{tima_aes}

As our ultimate experiment, we prepare an attack on a protected AES implementation. In this scenario, we target a 3-Share AES DOM hardware core. To ensure that random generator execution does not incur additional leakage, we consider an off-chip \textsc{TRNG}, which feeds in the masked operands into the FPGA target. As explained, we consider the attack in two scenarios. 
\subsubsection{Attacking Key Register Byte}\label{tima_aes_a}
Here, we perform the attack at the first clock-cycle where shares of the first key byte (as well as first input byte shares) are loaded into the target. Hence, regardless of the masked operations on the upcoming clocks, TIMA attacks the key (share) registers directly. We carry out the \textit{Profiling Stage} with $N=20,000$ number of traces to template masked key registers. Specifically, we execute \textit{TIMA} profiling for each bit of all key shares independently ($8\times3=24$ for each byte of the master key). It is also worth noting that shares are generated uniformly random and each trace is used to template for all target bits. More specifically, for each template target bit, we would have roughly $N_0=10,000$ traces where a target bit of the target share is \textbf{\texttt{0b0}}.

\begin{figure}[t]
  \begin{subfigure}[b]{0.49\columnwidth}
    \includegraphics[width=\linewidth]{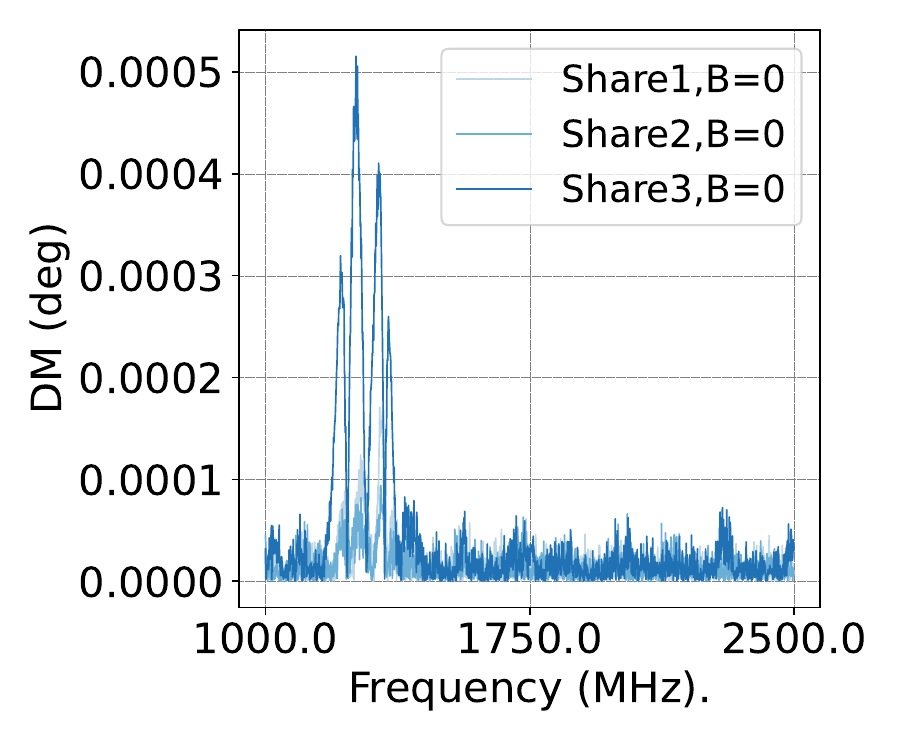}
    \caption{}
  \end{subfigure}
  \hfill 
  \begin{subfigure}[b]{0.49\columnwidth}
    \includegraphics[width=\linewidth]{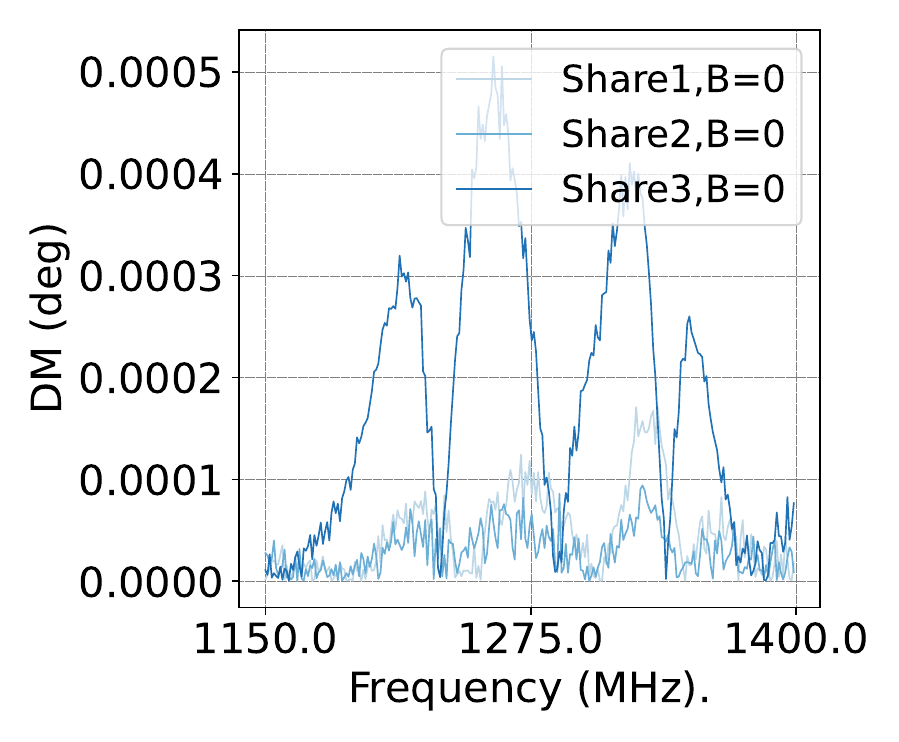}
    \caption{}
  \end{subfigure}
    \caption{Leakage measurement of \textit{Inter-Share} bits based on DM. (a) Wide frequency plot and (b) frequency zoomed-in plot.}
    \label{fig:temp_intra}
\end{figure}

\begin{figure}[h!]
  \begin{subfigure}[b]{0.49\columnwidth}
    \includegraphics[width=\linewidth]{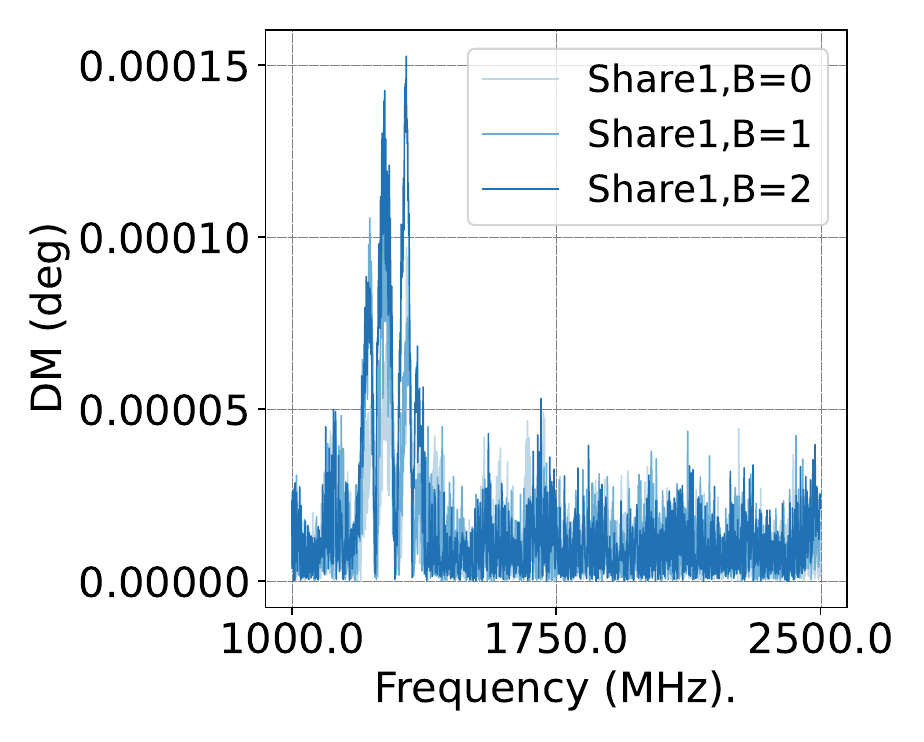}
  \end{subfigure}
  \hfill 
  \begin{subfigure}[b]{0.49\columnwidth}
    \includegraphics[width=\linewidth]{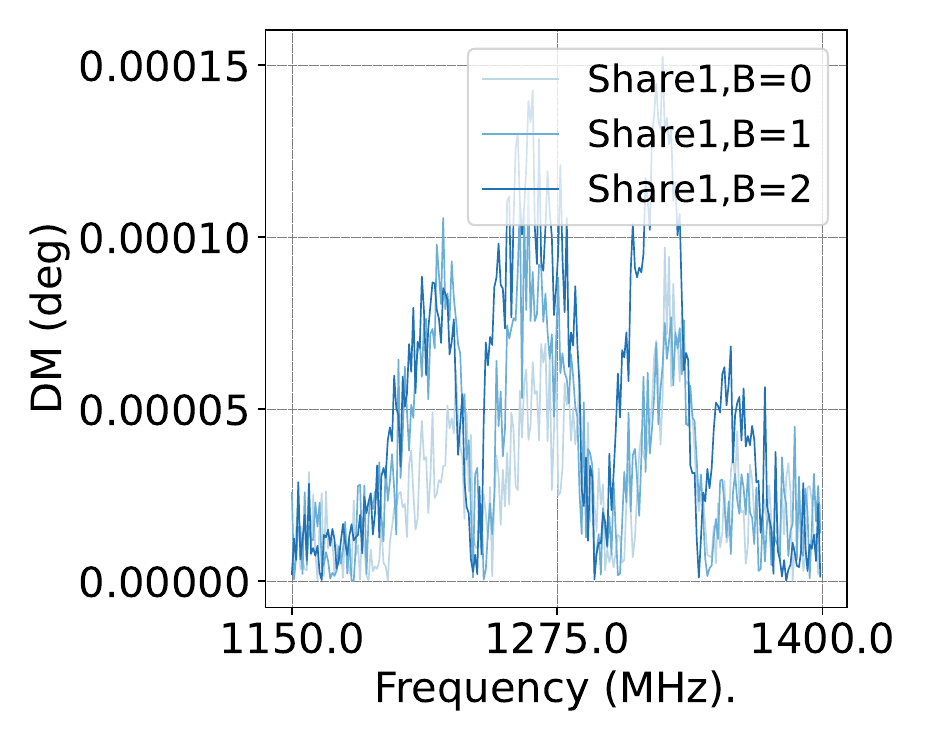}
  \end{subfigure}
    \caption{Leakage measurement of \textit{Intra-Share} bits based on DM. (a) Wide frequency plot and (b) frequency zoomed-in plot.}
    \label{fig:temp_inter}
\end{figure}

Figure~\ref{fig:temp_intra} and Figure\ref{fig:temp_inter} show the bit leakage model among bits on different shares and on the same share, respectively. Note that DM metric is shown here to highlight that distinguished POIs on different frequency points for each individual bit of each share enables \textit{TIMA} to effectively extract the master key.

\begin{figure}[t!]
  \centering \noindent
   \includegraphics[width=0.9\linewidth]{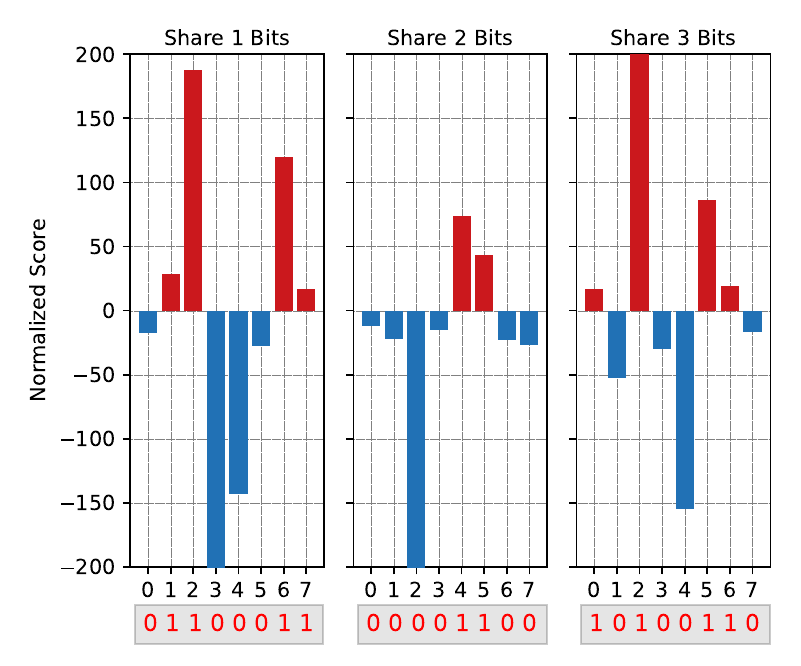}
	\caption{Extracting the bit values of all three shares for the first key byte, the first byte of master key can be computed as $K=S_1\oplus S_2 \oplus S_3= \texttt{0xC6} \oplus \texttt{0x30} \oplus \texttt{0x65}=\texttt{0x93}$}
            \label{fig:template_recover}                 
 \end{figure}

After \textit{Profiling Phase}, we perform a single trace attack on the DUT with unknown shares. In order to reduce the noise we carried out VNA-enabled averaging of $AV_{idx}=200$ on the the attack trace. \textit{TIMA} successfully recovers all the bits of each share individually. Figure~\ref{fig:template_recover} serves as an example to showcase the key extraction based on template scores of each target bit after attack stage. Combining extracted share bits could then be used to reconstruct the first byte of the master key as shown in Figure~\ref{fig:template_recover}.

  \begin{figure*}[t] 
\centering
\includegraphics[width=0.95\linewidth]{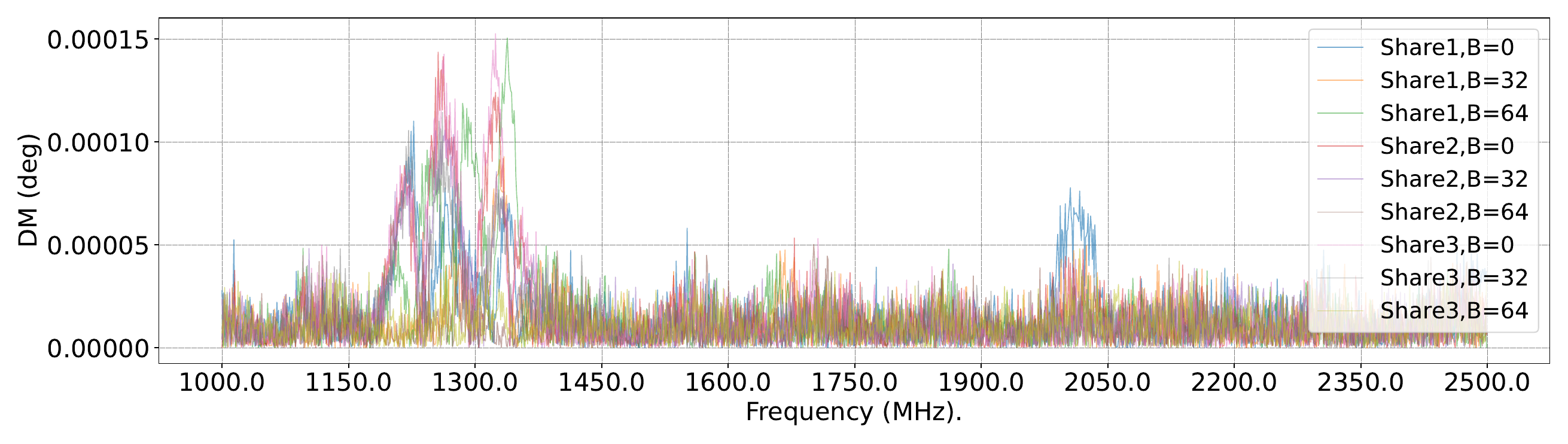}
\caption{illustration of full-length key leakage through frequency of AES-DOM using TIMA }
\label{fullkey}
\end{figure*}

\subsubsection{Analysing Full-length Key}\label{tima_aes_b}
We further focus on the key registers on the FPGA at the time before the AES-DOM is initiated. As random shares are stored on FPGA registers before loaded into the AES core (i.e., DOM core), we target the time-stamp where $aes\_start$ signal is set. Note that although DOM execution is byte-wise and shares are randomize in the key registers for each byte execution, original shares are loaded from other locations (additional wires and registers) into the core. These registers are targeted in this attack. We profile TIMA on full-length keys with 50,000 traces and follow the same procedure explained in the previous attack. Figure~\ref{fullkey} shows the DM metric over frequency domain for different bits on all three shares of the target key during TIMA \textit{Profiling Phase}. Note that similar to the previous experiment, DM is plotted for the averaged trace of approximately 25,000 traces for \textbf{\texttt{$b_t=0b0$}} and 25,000 traces for \textbf{\texttt{$b_t=0b1$}} ($DM_{b_t} = Abs(Mean(Tr|b_t=0b0)-Mean(Tr|b_t=0b1))$). As shown, each individual bit could visually be distinguished by DM metric.




\section{Discussions}\label{sec:dis}


{\subsection{Comparison with Power SCA}}
{There has been a numerous efforts to provide fair metrics \cite{papagiannopoulos2023side} and methodologies \cite{del2015side,bronchain2019leakage,masure2023information} to compare and analyze the leakage of various power analysis attacks. It is shown that time-series multivariate leakage in dynamic power attacks could be considered to improve attack success rate~\cite{grosso2015asca}. Impedance leakage is multivariate by nature but in contrast to dynamic leakage is time-constant. Although, dimension reduction techniques such as Principal Component Analysis (PCA) combined with leakage metrics such as Perceived Information (PI)~\cite{standaert2009unified} could be used to provide fair comparison with uni-variate leakage~\cite{del2015side} ( e.g. averaged static traces), further extensive experiments and modeling are required. }

{Moreover, the threat model assumptions, experimental conditions, and setup highly contributes to the SNR and success rate of the attack. Impedance attack presented in this work is somehow similar to static-side channels but differs in experimental details. Moradi~\cite{moradi2014side} and Moos~\cite{moos_static_2019} use climate chambers and full board clock halt (for noise reduction) to perform static SCA. Cassiers et al.~\cite{cassiers2023prime} removed the need for a climate chamber in a later work, but the attack still requires a low-noise amplifier, the control of the power supply and the source clock of the target, whereas impedance analysis utilizes only clock gating with no control over power supply or environment temperature. We believe that an accurate experimental comparison among static, dynamic, and impedance SCA should be under the same threat model, conditions, and target and deserves a future study. Another interesting course of research could be the leakage analysis of time-variant impedance profiles which could introduce time-frequency multivariate impedance attacks.}

\subsection{Possible Countermeasures}\label{sec:countermeasure}
As investigated throughout the paper, the nature of impedance leakage is directly caused by the physical placements of secret-dependant registers. 
Hence, a fundamental approach to resolve these leakages is real-time refreshing of the secret-dependant registers and/or routing. 
A series of hardware randomization solutions~\cite{guneysu2011generic,mentens2017hiding,koblah2022hardware} present methods including partial reconfiguration as \textit{Moving Target Defense}, to secure FPGA against SCA. Similar methodologies \cite{hettwer2019securing}, if deployed in an online manner, could be used to prevent impedance analysis. { Alternatively, system level approaches to detect clock-control on the device \cite{dumitru2023borrowed} could also be deployed as countermeasure to those impedance attacks which are performed in clock-controlled environment.}

\subsection{Clock control and impedance measurements}\label{sec:clock}
A clock controller is deployed in our measurement setup to ensure an accurate time stamp for our measurements. 
However, compared to static power side-channel where static snapshots without halted clocks could lead to an increase of the noise (due to additive noise caused by dynamic power consumption)~\cite{del2015side}, impedance analysis is not susceptible to time-varying measurements as long as the target intermediate values are stored (i.e., in Flip-Flops) in the circuit. 
This is mainly due to the unique and non-additive leakage of each individual (bit of) intermediate value through the frequency domain, which relaxes the clock control constraints in our threat model. 
On the other hand, as indicated by prior works~\cite{moos2019staticb,krachenfels2021real}, in many real-world cryptographic masked hardware and software implementations~\cite{vhdldom,ascadcode}, masked state/key registers are not overwritten every clock-cycle and are maintained for tens of clock cycles. Consequently, it is possible for the adversary to perform an iterative measurement without clock control to capture properly time-stamped measurements.

Furthermore, we stress that if target frequencies are known by the adversary (i.e., in the case of template attack), considering the typical VNA capabilities~\cite{keysightman}, frequency sweep could be done up-to two orders of magnitude faster, giving the adversary the advantage to perform multiple measurements during a single MHz clock execution of DUT.

\subsection{Effects of Wiring}
Previous researchers have shown that long wires in FPGA implementations could potentially increase the leakage for power side-channel attacks~\cite{giechaskiel2019leakier,giechaskiel2016information}. On the other hand, very close wiring between sensitive operands could also result in coupling effects~\cite{de2017does} that leak sensitive data. 
Hence, hardware routing and wiring should be done exceedingly carefully when it comes to implementing protected crypto-systems. Our experiments show that long wiring of secret dependant registers increases impedance leakages as well. 
Specifically, the deployment of long routing on the FPGA die realizes a series of buffers and intermediate elements that contributes to target impedance which could be exploited to reveal data. 


\section{Conclusion}\label{sec:Conclusion}

In this paper, we presented a novel non-invasive physical side-channel attack exploiting the data-dependent changes in the impedance of the chip.
We demonstrated that the temporarily stored contents in registers of an IC alter the die's impedance, which can be measured using scattering parameters commonly deployed in RF/microwave engineering.
Since registers at different locations affect the impedance profile of the system at various frequency bands, the content of these registers can be probed simultaneously by identifying the frequencies of interest.
To assess the threat of our discovered side-channel, we mounted several impedance analysis attacks against unprotected and protected hardware AES-128 implementations on an FPGA and showed that secret key bits could be recovered confidently.
In the case of the profiled attack, only a single trace was required to recover the secret key.
Our results challenged the effectiveness of masking schemes as side-channel countermeasures, and hence, we believe that the integration of specific hiding countermeasures in addition to masking is required to mitigate our attack.

\section*{Acknowledgement}\label{sec:Ack}
This effort was sponsored in part by NSF under the grant number 2150123 and in part by Electric Power Research Institute (EPRI).

\bibliographystyle{ACM-Reference-Format}

\bibliography{ref}




\end{document}